\documentclass[preprint]{aastex61}

\newcommand\aastex{AAS\TeX}

\usepackage{booktabs}
\usepackage{flushend} 
\newcommand{\VLSR}{V_{\rm LSR}}
\newcommand{\FeII}{[\ion{Fe}{2}]}
\newcommand{\SII}{[\ion{S}{2}]}

\newcommand{\kms}{km~s$^{-1}$}

\newcommand{\degree}{^{\circ}}
\newcommand{\Ha}{H${\alpha}$}
\newcommand{\Av}{A_{\rm V}}
\newcommand{\Ak}{A_{\rm K}}
\newcommand{\Ah}{A_{\rm H}}

\defcitealias{Oh2016a}{Paper I}

\received{xx}
\revised{xx}
\accepted{xx}

\shorttitle{\aastex\ draft}
\shortauthors{Oh et al.}

\begin{document}

\title{High-resolution near-IR Spectral mapping with H$_{2}$ and {\FeII} lines of Multiple Outflows around LkH$\alpha$ 234}

\correspondingauthor{Heeyoung Oh}
\email{hyoh@kasi.re.kr}

\author[0000-0002-0418-5335]{Heeyoung Oh}
\affil{Department of Physics and Astronomy, Seoul National University, 1 Gwanak-ro, Gwanak-gu, Seoul 08826, Korea}
\affiliation{Korea Astronomy and Space Science Institute, 776 Daedeok-daero, Yuseong-gu, Daejeon 34055, Korea}
\affil{Department of Astronomy, University of Texas at Austin, Austin, TX 78712, USA}

\author{Tae-Soo Pyo}
\affiliation{Subaru Telescope, National Astronomical Observatory of Japan, National Institutes of Natural Sciences (NINS), 650 North A'ohoku Place, Hilo, HI 96720, USA}
\affiliation{School of Mathematical and Physical Science, SOKENDAI (The Graduate University for Advanced Studies), Hayama, Kanagawa 240-0193, Japan}

\author{Bon-Chul Koo}
\affiliation{Department of Physics and Astronomy, Seoul National University, 1 Gwanak-ro, Gwanak-gu, Seoul 08826, Korea}

\author{In-Soo Yuk}
\affiliation{Korea Astronomy and Space Science Institute, 776 Daedeok-daero, Yuseong-gu, Daejeon 34055, Korea}

\author{Kyle F. Kaplan}
\affil{Department of Astronomy, University of Texas at Austin, Austin, TX 78712, USA}
\affil{Department of Astronomy/Steward Observatory, The University of Arizona, 933 North Cherry Avenue, Tucson, AZ 85721}

\author{Yong-Hyun Lee}
\affil{Department of Physics and Astronomy, Seoul National University, 1 Gwanak-ro, Gwanak-gu, Seoul 08826, Korea}

\author{Kimberly R. Sokal}
\affil{Department of Astronomy, University of Texas at Austin, Austin, TX 78712, USA}

\author{Gregory N. Mace}
\affil{Department of Astronomy, University of Texas at Austin, Austin, TX 78712, USA}

\author{Chan Park}
\affiliation{Korea Astronomy and Space Science Institute, 776 Daedeok-daero, Yuseong-gu, Daejeon 34055, Korea}

\author{Jae-Joon Lee}
\affiliation{Korea Astronomy and Space Science Institute, 776 Daedeok-daero, Yuseong-gu, Daejeon 34055, Korea}

\author{Byeong-Gon Park}
\affiliation{Korea Astronomy and Space Science Institute, 776 Daedeok-daero, Yuseong-gu, Daejeon 34055, Korea}
\affiliation{Korea University of Science and Technology, 217 Gajeong-ro, Yuseong-gu, Daejeon 34113, Korea}

\author{Narae Hwang}
\affiliation{Korea Astronomy and Space Science Institute, 776 Daedeok-daero, Yuseong-gu, Daejeon 34055, Korea}

\author{Hwihyun Kim}
\affiliation{Gemini Observatory, Casilla 603, La Serena, Chile}

\author{Daniel T. Jaffe}
\affil{Department of Astronomy, University of Texas at Austin, Austin, TX 78712, USA}




\begin{abstract}

We present a high-resolution, near-IR spectroscopic study of multiple outflows in the LkH$\alpha$ 234 star formation region using the Immersion GRating INfrared Spectrometer (IGRINS). Spectral mapping over the blueshifted emission of HH 167 allowed us to distinguish at least three separate, spatially overlapped, outflows in H${_2}$ and {\FeII} emission. We show that the H${_2}$ emission represents not a single jet, but complex multiple outflows driven by three known embedded sources: MM1, VLA 2, and VLA 3.
There is a redshifted H${_2}$ outflow at a low velocity, $\VLSR$ $<$ $+$50 {\kms}, with respect to the systemic velocity of $\VLSR$ $=$ $-$11.5 {\kms}, that coincides with the H${_2}$O masers seen in earlier radio observations two arcseconds southwest of VLA 2.
We found that the previously detected {\FeII} jet with $|$$\VLSR$$|$ $>$ 100 {\kms} driven by VLA 3B is also detected in H${_2}$ emission, and confirm that  this jet has a position angle about 240$\degree$. 
Spectra of the redshifted knots at 14$\arcsec$$-$65$\arcsec$ northeast of LkH$\alpha$ 234 are presented for the first time. These spectra also provide clues to the existence of multiple outflows.
We detected high-velocity (50$-$120 {\kms}) H${_2}$ gas in the multiple outflows around LkH$\alpha$ 234. Since these gases move at speeds well over the dissociation velocity ($>$ 40 {\kms}), the emission must originate from the jet itself rather than H${_2}$ gas in the ambient medium. 
Also, position-velocity diagrams and excitation diagram indicate that emission from knot C in HH 167 come from two different phenomena, shocks and photodissociation.

\end{abstract}




\section{Introduction} \label{sec:intro}

Mass accretion and outflows are essential processes in the formation of stars and planets, as they remove the angular momentum of the infalling material \citep{Hartigan1995}.
Massive star formation is not understood well \citep{Zinnecker2007}, but massive stars are thought to form in multiple systems and in filamentary structures in molecular clouds \citep[e.g.,][]{Pineda2015}.
To understand the early stages of the formation of multiple systems, it is important to probe the alignment and orientation of disks and outflows in multiple protostars, in testing competing theories on the massive star formation, such as ``competitive accretion'' \citep{Bonnell1997,Bonnell2001} and ``stellar collisions and mergers'' \citep{Bonnell1998,Zinnecker2007}.


The intermediate-mass star Lk{\Ha} 234 \citep[$\sim$ 8.5 M$_{\sun}$,][]{Hillenbrand1992} is located in the NGC 7129 cluster at a distance of $\sim$ 1.25 kpc \citep{Shevchenko1989}. Lk{\Ha} 234 and the surrounding star forming region are a good place to study the nature of young multiple systems, because this region is one of the most complicated star forming regions, with multiple outflows from different protostars \citep[e.g.,][]{Kato2011,Oh2016a}.
Figure \ref{fig:slitposition} and \ref{fig:source_positions}(a) show the positions of the Young Stellar Object (YSO) candidates and axes of the multiple outflows around LkH$\alpha$ 234. The positions of sources are taken from the radio study by \citet{Trinidad2004}.

The first outflow in this region was identified by \citet{Edwards1983} in J = 1$-$0 $^{12}$CO, with a redshifted velocity of $\sim$ 10 {\kms} to the northeast of Lk{\Ha} 234 with a scale of $\sim$ 4$\arcmin$. 
Herbig-Haro (HH) 167 is an optical jet discovered by \citet{Ray1990}, and they found high-velocity {\SII} emission with a velocity over $-$100 {\kms} extending more than 30$\arcsec$ with a  position angle (P.A.) $\sim$ 252$\degree$. \citet{Ray1990} suggested that this jet could be the counterpart of the red CO lobe, and the prominent {\SII} emission within $\sim$ 15$\arcsec$ of Lk{\Ha} 234 are named knots A, B, and C. 
In the near-IR observation, H$_{2}$ emission observed by \citet{Schultz1995} and \citet{Cabrit1997} is consistent with the {\SII} knots A, B, and C in HH 167.
In larger scale, \citet{Eisloffel2000} detected shocked H$_{2}$ emission in the CO outflow region in \citet{Edwards1983}.
\citet{McGroarty2004} found that the {\SII} emission is spread over 22$\arcmin$ on the sky, indicating a parsec-scale jet.

Radio, millimeter, and mid-IR observations have suggested the presence of YSO to the immediate northwest of Lk{\Ha} 234.
The radio continuum sources VLA 1, 2, and 3 are detected at $\sim$ 6$\arcsec$, 3$\arcsec$, and 2$\arcsec$ from Lk{\Ha} 234 \citep{Trinidad2004}, respectively. The strongest source VLA 3 is a binary with components of 3A and 3B, and both sources are thought to be emanating radio thermal jets \citep{Trinidad2004}. Another embedded source FIRS1-MM1 was discoverd in millimeter observations by \citet{Fuente2001}, $\sim$ 4$\arcsec$ northwest of Lk{\Ha} 234.
A 10 $\micron$ source (IRS 6) is spatially associated with VLA 3 \citep{Cabrit1997}. \citet{Kato2011} identified mid-IR sources NW 1 and 2, at the positions of IRS 6 and VLA 2, respectively.
In \citet{Oh2016a} (hereafter \citetalias{Oh2016a}), we presented more details on detections of multiple sources and outflows around Lk{\Ha} 234.

\citet{Trinidad2004} suggested that VLA 1 is associated with a radio jet on the basis of its elongated morphology.
The near-IR H$_{2}$ jet has major axis with a P.A. $=$ 227$\degree$, being aligned with FIRS1-MM1 \citep{Fuente2001}.
\citet{Trinidad2004} and \citet{Torrelles2014} showed that the jet has bipolar kinematics centered on VLA 2, determined from the proper motions of H$_{2}$O masers. They suggest it is possibly related to the optical {\SII} outflow based on their similarity in axis, although the origin of {\SII} outflow is not yet clear.
The P.A. of radio thermal jet around VLA 3B is $\sim$ 230 $\degree$, and in \citetalias{Oh2016a} we found that {\FeII} emission arises from the outflow driven by VLA 3B.
None of the outflows seem associated with Lk{\Ha} 234, implying that this YSO could be in a later evolutionary stage than others \citep{Girart2016}.

With the millimeter observation, \citet{Girart2016} revealed the filamentary dust structure surrounding VLA 1$-$3 and FIRS1-MM1.They indicated the possibility of a sequential star formation within the filaments based on evolutional stages. They also detected a compact SiO bipolar outflow with an axis close to the plane of the sky which could be a counterpart of H$_{2}$O masers.

Spectral mapping using near-IR, high-resolution spectroscopy is a powerful tool for analysis of the nature of multiple outflows \citep[e.g.,][]{Oh2016b,Youngblood2016}. The channel maps made from a three-dimensional (3D; x, y, and velocity) datacube provides high contrast images with ultra narrow-band widths ($\Delta$$v$ $\sim$ 10 {\kms}). This is better than narrow-band imaging filters that have widths corresponding to several thousand {\kms} in velocity.
There have been several big H$_{2}$ surveys such as \citet{Froebrich2011} and \citet{Walawender2013}, but spectroscopy is the only tool available to measure the amount of shocked material in a definitive way using ratios between multiple emission lines, as well as the kinematic information.

In this work we extensively study the region around LkH$\alpha$ 234 with a six pointing spectral map obtained with the Immersion GRating INfrared Spectrometer \citep[IGRINS;][]{Yuk2010,Park2014,Mace2016}.
\citetalias{Oh2016a} reported an interesting result revealing a new jet driven by VLA 3B with the high-velocity {\FeII} emission, but the narrow spatial coverage and the misalignment of the slit with the jet position angles limited the detailed study of the multiple outflow structures.
In this paper, we present the result from a spectral map with wider coverage, which allows us to probe the overall kinematics and physics of HH 167 using the mapping with full coverage over the near-IR H$_{2}$ jet including knot C.
We construct a datacube of H$_{2}$ and {\FeII} emission lines, and the channel maps of the datacube allow us to distinguish multiple outflows overlapping spatially but separated in velocity and position angle. We also discuss the orientation of multiple outflows not known before. In addition to the spectral mapping, we present the dynamics and shock properties using the first spectra obtained at the position of the redshifed CO lobe at northeast of HH 167 and surrounding photodissociation region (PDR).
Finally, we discuss the role of multiple outflows from a small cluster ridding themselves of their dense envelopes, and adding turbulence to the clouds that help to support them in the large scale.

\section{Observation and Data Reduction} \label{sec:obs}

\subsection{NIR Imaging Data}
In order to find appropriate slit positions for the observations and to obtain spatial information around LkH$\alpha$ 234,
we used a H$_{2}$ 1$-$0 S(1) narrow-band image of NGC 7129 obtained by CFHT-IR \citep{Starr2000} at the 3.6m Canada-France-Hawaii Telescope (CFHT). The data is obtained on 2003 November 11 (UT).
CFHT-IR camera used the Rockwell HAWAII 1k $\times$ 1k HgCdTe array,which provides a field of view of 3$\farcm$6 $\times$ 3$\farcm$6 with a pixel scale of 0$\farcs$211 pixel$^{-1}$.
The data was obtained using a narrow-band filter (H$_{2}$ 1$-$0 S(1), $\#$5339) with the center wavelength and full width at half maximum (FWHM) of $\lambda$$_{c}$ $=$ 2.122 $\micron$ and $\Delta\lambda$ $=$ 200 {\AA}, respectively.
We also used $K$-continuum filter ($\#$5342) image, with $\lambda$$_{c}$ $=$ 2.260 $\micron$ and $\Delta\lambda$ $=$ 600 {\AA}.

The telescope was dithered to four different directions: northeast, northwest, southeast, and southwest, with respect to the center position (J2000 $=$ 21:43:06, $+$66:07:09) by 30$\arcsec$.
The total on-source exposure times were 18 s for both H$_{2}$ and $K$-continuum filters, with each single exposure of 2 s. The median seeing during the observation in $K$-band was $\sim$ 1$\arcsec$.

All the raw data were downloaded from 
Canadian Astronomy Data Centre\footnote{\url{http://www.cadc-ccda.hia-iha.nrc-cnrc.gc.ca/}},
and we conducted a data reduction using standard technique for near-IR imaging data.
First, all the science frames were divided by a normalized flat frame,
and then, were subtracted by a sky frame
which was derived from the median-averaging of all the flat-fielded frames.
The astrometry was corrected by comparing the bright, isolated stars in the field with
Two Micron All Sky Survey (2MASS) Point Source Catalog \citep[PSC;][]{Skrutskie2006}.
All the pre-processed frames were then combined into a final image.

\subsection{IGRINS Observation}
We used data from IGRINS at two different telescopes, which are the 2.7m Harlan J. Smith Telescope (HJST) at the McDonald Observatory and the 4.3m Discovery Channel Telescope (DCT) at Lowell Observatory.
IGRINS covers the whole wavelength range of the infrared $H$- and $K$-bands (1.49$-$2.46 $\micron$) simultaneously, with a spectral resolving power $R$ $\equiv$ $\lambda/\Delta\lambda$ $\sim$ 45,000. The wavelength coverage and the resolving power of IGRINS are the same on both telescopes. The resolving power corresponds to a velocity resolution ($\Delta v$) of 7 {\kms}, with $\sim$ 3.5 pixel sampling.
Table \ref{tbl:obs_log} presents the summary of the IGRINS observations, and the details of observations using HJST and DCT are described below.

\subsubsection{Spectral Mapping with HJST}\label{sec:obs_mapping}
On 2015 August 6 (UT), we obtained a spectral map toward the H${_2}$ jet (HH 167) southwest of LkH$\alpha$ 234 using HJST. The slit size on the sky, which changes depending on the telescope, was 1\farcs0 $\times$ 15\farcs0 at HJST. 
The pixel scale is 0\farcs24 $-$ 0\farcs29 pixel$^{-1}$ along the slit, and the value is larger in higher orders. Auto-guiding is performed during each exposure with a $K$-band slit-viewing camera (pixel scale = 0\farcs12 pixel$^{-1}$). 
The guiding uncertainty was smaller than 0\farcs4 on average and the FWHM of stars in the $K$-band slit-viewing camera was $\sim$0\farcs9, which establishes the angular resolution of the spectral map.

\begin{figure}
\epsscale{1.1}
\plotone{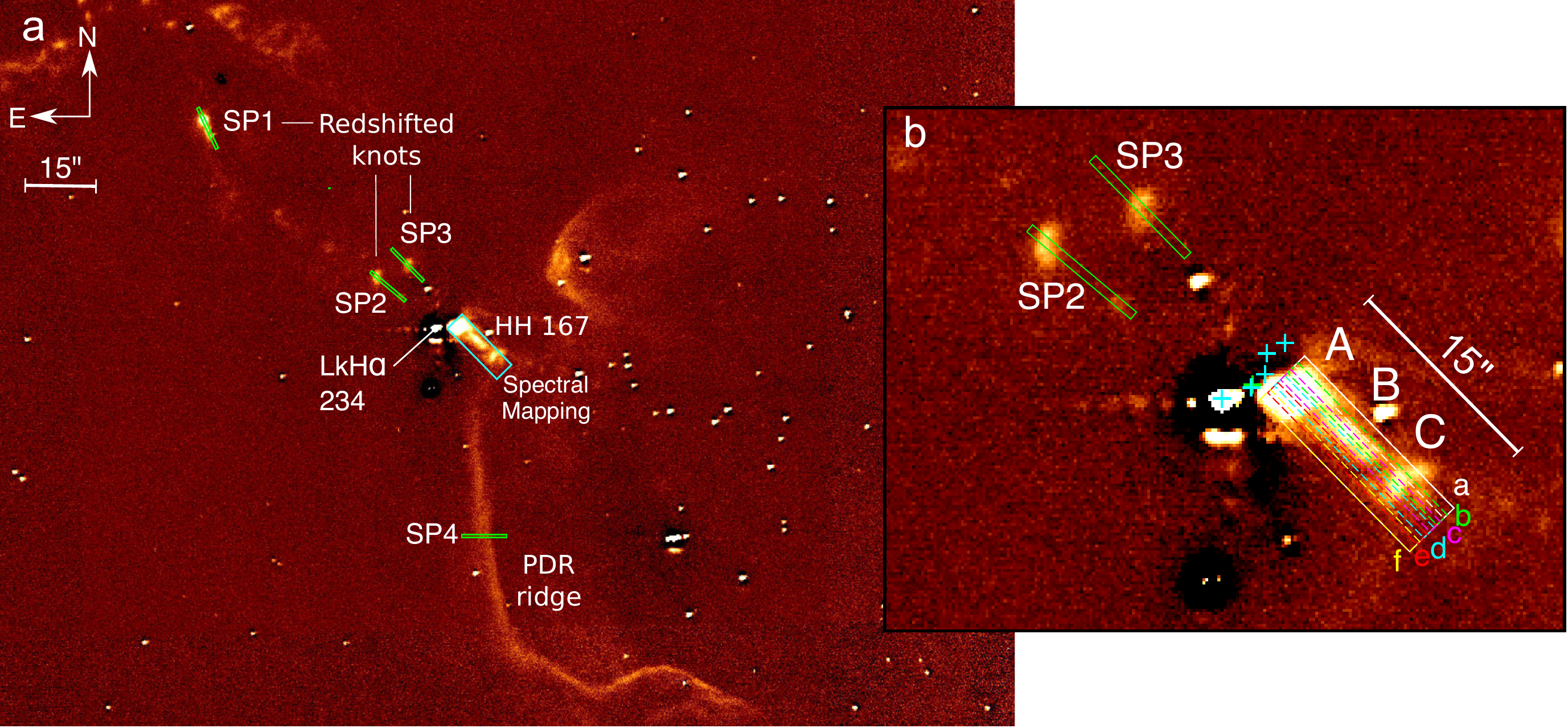}
\caption{IGRINS slit positions for the spectral map of HH 167, observed at the HJST, and of SP1$-$4 taken with IGRINS on the DCT. Slit sizes are 1\farcs0 (W) $\times$ 15\farcs0 (L) and 0\farcs63 (W) $\times$ 9\farcs3 (L) at HJST and DCT, respectively. (a) The background image is a continuum-subtracted H$_{2}$ 1$-$0 S(1) $\lambda$2.122 $\micron$ narrow-band image of the LkH$\alpha$ 234 star forming region obtained by CFHT. The positive-negative spots in the image are the residuals from the continuum subtraction. The spectral mapping area is indicated as a cyan rectangle along the H$_{2}$ jet (HH 167).
SP1$-$3 correspond to the redshifted knots northeast of LkH$\alpha$ 234 and SP4 covers part of the `PDR ridge'. (b) Spectral mapping at six different positions from `a' to `f'. Knot A, B, and C of HH 167 is labelled. The crosses mark the sources LkH$\alpha$ 234 (J2000 $=$ 21:43:06.816, $+$66:06:54.26), VLA 3 (A and B), VLA 2, MM1, and VLA 1 from the southeast to the northwest.  \label{fig:slitposition}}
\end{figure}

The spectral map was completed by performing six adjacent observations, offset by 0\farcs7 steps perpendicular to the slit length. This map coveres a $\sim$ 15$\arcsec$ $\times$ 4\farcs3 area including knot A, B, and C of HH 167 jet. Figure \ref{fig:slitposition} shows the slit positions on the sky. The slit position angle (P.A.) was 225$\degree$. The total on-source integration time was 600 s at each slit position. The observing sequence for this program included $object-sky-object$ observations for each slit position. The sky frame was obtained at a position 180$\arcsec$ east of the on-source position. We also observed an A0V-type star (HR 8598, $K$=6.32mag) for telluric correction and Th-Ar and halogen lamp frames were taken for wavelength calibration and flat-fielding, respectively.

On 2017 June 9 (UT), we acquired additional spectra in order to probe the properties of high-velocity H$_{2}$ emission with the line ratios. For the high excitation lines, higher signal-to-noise (S/N) data are required than that obtained with the original mapping.
A deep pointing at single position `b' (see Figure \ref{fig:slitposition}), where we found the strongest high-velocity H$_{2}$ emission, was conducted. The telescope and instrument settings, and slit P.A. were kept consistent with those used in previous observation in 2015. The total on-source exposure time for this single pointing was 2400 s and an A0V-type star (HD 191940, K$=$6.60 mag) was observed as a telluric star.

\subsubsection{Single Slit Positions with DCT}
In addition to the spectral mapping over HH 167, we obtained spectra at additional slit positions of interest. The data were obtained on 2016 November 19$-$20 (UT), while IGRINS was installed at the DCT. On the DCT, the slit size was 0\farcs63 $\times$ 9\farcs3. The pixel scale along the slit was therefore reduced by a factor of $\sim$ 0.63. The wavelength coverage and spectral resolving power of IGRINS remained the same.
We selected 4 separate slit positions on the H${_2}$ emission; three positions on the northeast of LkH$\alpha$ 234, which are spatially coincident with the redshifted CO outflow \citep{Edwards1983,Eisloffel2000}, and one position on the ``PDR ridge'' \citep{Morris2004} to the south of LkH$\alpha$ 234. The slit positions (SP1$-$4) are shown in Figure \ref{fig:slitposition}. The slit P.A. was 25$\degree$, 50$\degree$, 44\fdg5, and 90$\degree$ for SP1$-$4, respectively. An A0V-type telluric star (HR 8598, K=6.32 mag) was also observed. The total integration time was 300 s for SP1 and SP4 and 600 s for SP2 and SP3, respectively. 

\subsubsection{Data Reduction}
Basic data reduction was done using the IGRINS Pipeline Package v2.2.0-alpha.3\footnote{The IGRINS Pipeline Package is downloadable at \url{https://github.com/igrins/plp}. (doi:10.5281/zenodo.18579).} \citep[PLP,][]{Lee2017}. The PLP performs sky subtraction, flat-fielding, bad-pixel correction, aperture extraction, and wavelength calibration. After running the PLP, additional processes are conducted using Plotspec\footnote{https://github.com/kfkaplan/plotspec} \citep{Kaplan2017}, which has been developed for the processing of two-dimensional (2D) spectra from IGRINS data. Plotspec provides continuous 2D spectra of all the IGRINS $H$- and $K$-band orders, removal of stellar photospheric absorption lines from the standard star, telluric correction, and relative flux calibration. Continuum is subtracted using pixel values obtained by a robust-median filter running along the wavelength direction.
With the Plotspec code we then constructed a 3D datacube from the spectral mapping data. 

\section{Results} \label{sec:result}
We detected 14 H$_{2}$ and 5 {\FeII} lines from knots A, B, and C of HH 167 in the 1.49$-$2.46 $\micron$ range covered by IGRINS. The lists of the detected lines are reported in the Tables 1 and 2 of \citetalias{Oh2016a}.
From the redshifted knots SP1$-$3, we detected 8$-$11 H$_{2}$ lines and we do not detect any {\FeII} emission, probably due to the large extinction in the $H$-band.
In SP4, which covers the PDR ridge (see Figure \ref{fig:slitposition}), high excitation lines of H$_{2}$ with $v$ (the upper vibrational level of H$_{2}$) $\leq$ 9 are detected and {\FeII} emission were found. Table \ref{tbl:pdr} lists the detected H$_{2}$ lines from SP1$-$SP4.
Our analysis of the datacube constructed from the spectral map utilizes channel maps and position velocity diagrams (PVDs).
The strongest H$_{2}$ 1$-$0 S(1) 2.122 $\micron$ and
{\FeII} $a^{4}D_{7/2} - a^{4}F_{9/2}$ $\lambda$1.644 $\micron$ lines are used to study the kinematics and the origin of the multiple outflows. 
We presented the detailed physics from the line ratios between different {\FeII} lines in \citetalias{Oh2016a}, by estimating an electron density of $\sim$ 1.1 $\times$ 10$^{4}$ cm$^{-3}$ which is similar to or slightly smaller than the values in outflows from T-Tauri stars or Class 0$-$I sources.

\subsection{Spectral Mapping of the HH167 Outflow}
\subsubsection{Monochromatic images} \label{sec:monochromatic}
\begin{figure}
\epsscale{.9}
\plotone{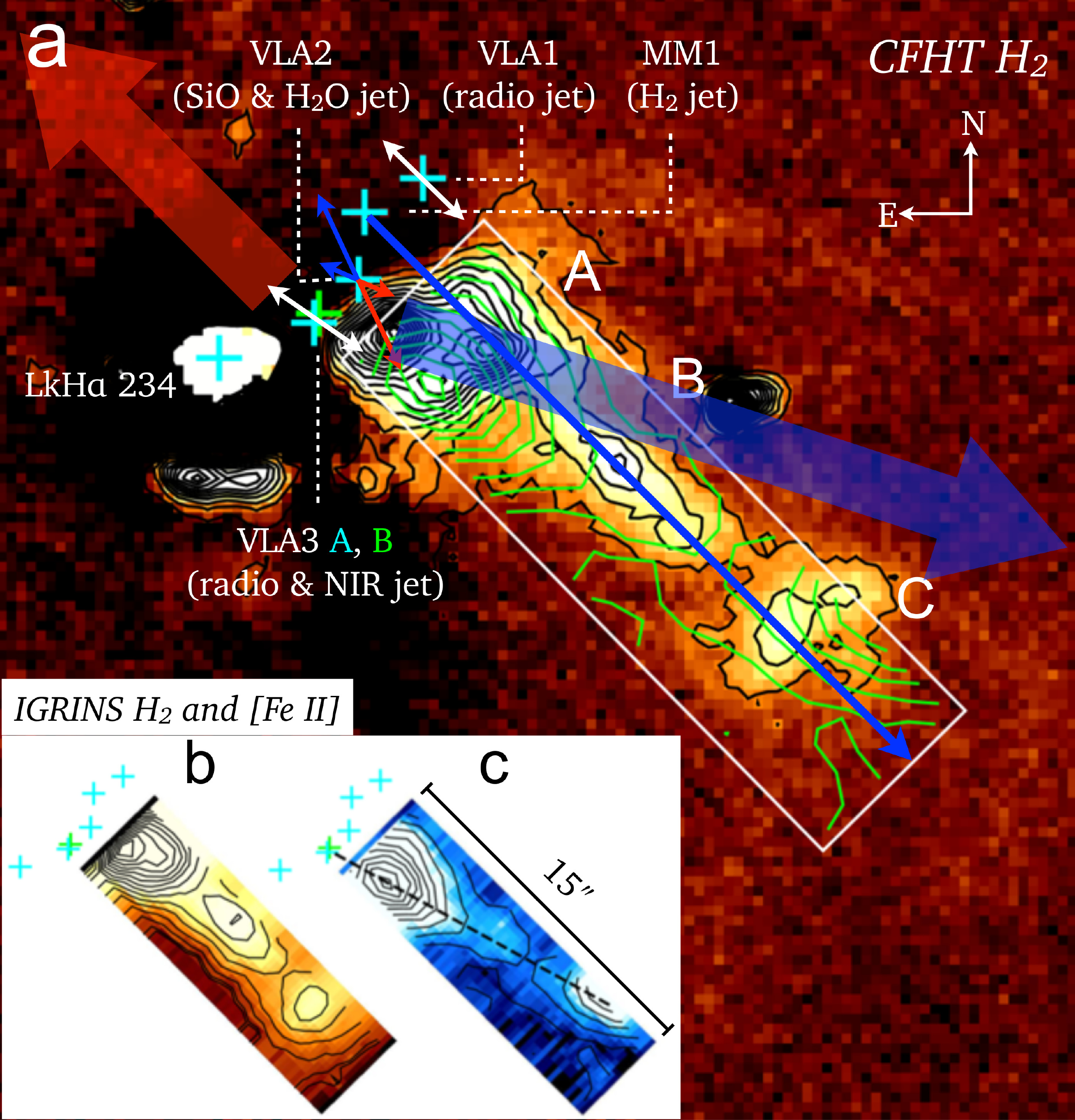}
\caption{(a) The positions of the YSOs and the axes of multiple outflows around the LkH$\alpha$ 234 region. The background image is a continuum-subtracted H$_{2}$ 1$-$0 S(1) narrow-band image and the black contours indicate its intensity in logarithmic scale. The size of the map is 15$\arcsec$ $\times$ 4\farcs3, a rectangle drawn with a white solid line. The residuals after subtracting LkH$\alpha$ 234 leave a small white area, and the region around LkH$\alpha$ 234 is affected by artificial features due to continuum subtraction. (b)$-$(c) Monochromatic line images of H$_{2}$ 1$-$0 S(1) $\lambda$2.122 $\micron$ and {\FeII} $\lambda$1.644 $\mu$m lines constructed from IGRINS spectral mapping, respectively. The black contours represent intensity levels of each emission line. H$_{2}$ and {\FeII} contours start from 3$\sigma$ and 2$\sigma$ levels, and increase with equal intervals in square root scale to highest levels of 230$\sigma$ and 80$\sigma$, respectively. The contours of {\FeII} $\lambda$1.644 $\mu$m emission are superposed on (a) in green. The crosses mark the sources LkH$\alpha$ 234 (J2000 $=$ 21:43:06.816, $+$66:06:54.26), VLA 3A, VLA 3B, VLA 2, MM1, and VLA 1 from the southeast to the northwest. Thin arrows around the sources show the axes of the outflows. Thick blue and red arrows indicate the axes of the blueshifted optical {\SII} jet with P.A. $\sim$ 252$\degree$ and the redshifted CO lobe, respectively. The black dashed line in (c) is the axis of {\FeII} emission with P.A. $=$ 240$\degree$. \label{fig:source_positions}}
\end{figure}

Figure \ref{fig:source_positions}(a) shows a close view of the narrow-band H$_{2}$ 1$-$0 S(1) emission image taken with CFHT-IR/CFHT shown in Figure 1. 
Knot A shows complicated structure with multiple peaks in intensity. It consists of two bright peaks at eastern and western parts of knot A, and another two minor peaks at the south and north. We note the possible contamination in H$_{2}$ emission close to the sources due to the residuals after the subtraction of continuum emission from LkH$\alpha$ 234.
In Section \ref{sec:channel}, we will show that these different peaks in knot A trace different velocities in the channel maps. The shape of knot B is elongated along the northeast-southwest direction with a size of 4\farcs5 in length, showing a good agreement with the major axis of P.A. $\sim$ 225$\degree$. Knot C is located at the southwest tip of the stream and shows clumpy structure.

The spectral mapping area is overlaid on Figure \ref{fig:source_positions}(a) with a white rectangle. 
Figure \ref{fig:source_positions}(b) and \ref{fig:source_positions}(c) show the images of H$_{2}$ 1$-$0 S(1) $\lambda$2.122 $\micron$ and {\FeII} $\lambda$1.644 $\micron$ emission lines integrated over a velocity range of $-$180 {\kms} $<$ ${\VLSR}$ $<$ $+$60 {\kms}. In H$_{2}$ 1$-$0 S(1) emission, the integrated intensity distribution matches well with the H$_{2}$ narrow-band image, showing the substructures of knots A, B, and C. The P.A. of H$_{2}$ emission is $\sim$ 225$\degree$, as shown in the previous imaging study by \citet{Cabrit1997}. The peak intensity is highest in knot A, and it is weaker at knot B and C by fractions of 0.25 and 0.16, respectively. 

{\FeII} emission shown in Figure \ref{fig:source_positions}(c) is very different from H$_{2}$ in morphology.
The axis of {\FeII} emission is $\sim$ 240$\degree$, and is well aligned with the position of radio continuum source VLA 3. This result supports the idea presented in \citetalias{Oh2016a}, which argued that this {\FeII} emission arises from the outflow driven by VLA 3B. The details of this emission will be discussed more in Section \ref{sec:discussion} in combination with the results from channel maps.
{\FeII} emission shows a single peak in knot A and its peak position is consistent with a small peak in H$_{2}$ emission in the southern part of knot A.  
The {\FeII} emission peak in knot A is located 1$\farcs$5 south from the H$_{2}$ peak in Figure \ref{fig:source_positions}(b).
The peak {\FeII} intensity of knot C is $\sim$ 0.34 times the knot A intensity, although knot A and C are not fully covered in the slit-scan area.
Emission from knot B is about 10 times weaker than knot A. 
In knot C the {\FeII} emission peaks 1$\farcs$5 to the west of the H$_{2}$ peak emission.
We note that extinction can cause the differences in morphology of {\FeII} and H$_{2}$ emission, as extinction in the $H$-band is larger than in $K$-band with a ratio of $\Ah$/$\Ak$ $\sim$ 1.56 \citep{Rieke1985}. 
The measurement of the extinction is described in Section \ref{sec:Av} in detail.

\subsubsection{Channel maps} \label{sec:channel}

Figure \ref{fig:channel} shows channel maps of the H$_{2}$ $\lambda$2.122 $\micron$ 1$-$0 S(1) and {\FeII} $\lambda$1.644 $\micron$ emission lines. The channel maps are constructed with 20 {\kms} velocity intervals from $-$180 {\kms} $<$ ${\VLSR}$ $<$ $+$60 {\kms}. 
We take the systemic velocity as $\VLSR$ $\sim$ $-$11.5 {\kms}, which is the central velocity of the molecular emission (e.g., SO$_{2}$) from radio continuum source VLA 2 taken from \citet{Girart2016}. 
The channel maps show that the H$_{2}$ and {\FeII} emission are different not only in position, but also in velocity. H$_{2}$ is dominant in lower velocity when compared to {\FeII} emission, but it shows more complex structure with multiple velocity components detected over a very wide velocity range.

In the channel maps of $-$30, $+$10, and $-$10 {\kms}, we mark the peaks corresponding to `A1$\arcmin$', `A2$\arcmin$', and `B' identified in \citetalias{Oh2016a}. We also mark a new peak `C', which arises from knot C. 
At the position of knot A, H$_{2}$ is detected in all velocity channels, including both blueshifted and redshifted components.
The two strong velocity peaks A1$\arcmin$ and A2$\arcmin$ show positional agreement with two strong intensity peaks in knot A shown in Figure \ref{fig:source_positions}(a). 
Peak A1$\arcmin$ shows the highest intensity level among the area of spectral mapping. 
In the channel centered on $-$10 {\kms}, knot A shows double peaks with somewhat `transitional' morphology between A1$\arcmin$ and A2$\arcmin$.
The emission from knot B is detected in the range of $-$60 {\kms} $<$ ${\VLSR}$ $<$ $+$20 {\kms}, with a peak value near the systemic velocity of $-$11.5 {\kms}. Knot C is more prominent at higher velocity than both knot A and B, and is centered on the $-$70 {\kms} channel map.

Channel maps of {\FeII} line show knot A has the strongest emission at $-$110 {\kms}. The peak of {\FeII} emission from knot B is not clear, but is most prominent in the channel of $-$130 {\kms}. {\FeII} emission from knot C peaks at the velocity higher than that of knot A and B, similar to the H$_{2}$ emission. The {\FeII} peaks at much higher velocity than H$_{2}$ in all of the knots.

In the channel maps, the dotted lines and circles in different colors mark the axes and the positions of the features which might arise from the different outflows. 
The magenta lines in the maps of H$_{2}$ and {\FeII} mark the P.A. of 240$\degree$. These lines correspond to the black dashed line in Figure \ref{fig:source_positions}(c), which is aligned well with the radio continuum VLA 3.
This axis is traced in all three knots in both H$_{2}$ and {\FeII}, in wide velocity range of $-$180 {\kms} $<$ ${\VLSR}$ $<$ $+$60 {\kms}. The emission along this axis is brighter in {\FeII} than H$_{2}$, while weak H$_{2}$ traces the whole {\FeII} emission along the magenta line, as clearly seen at ${\VLSR}$ $=$ $-$130 {\kms}. Along this axis, the peak velocities of both H$_{2}$ and {\FeII} emission are very similar in all knots, as shown in channels of $-$110 {\kms}, $-$130 {\kms}, and $-$150 {\kms} for knot A, B and C, respectively.
The green vertical line shows the axis of P.A. $=$ 225$\degree$, which is equal to the major axis of H$_{2}$ emission in Figure \ref{fig:source_positions}(a). Along this axis, the H$_{2}$ emission is again detected in all three knots, in wide velocity range of $-$100 {\kms} $<$ ${\VLSR}$ $<$ $+$20 {\kms}. Very weak {\FeII} emission from knot B and C trace this axis, but is marginally detected with $\sim$ 1$\sigma$ level (see panels (b) and (c) in Figure \ref{fig:pvd}).
Black ellipses in the channel maps of $+$10 and $+$30 {\kms} mark the redshifted component in knot A position. 
Additionally, blue circles in channel maps at $-$50 and $-$30 {\kms} indicate emission from peak A1$\arcmin$. 

\subsubsection{Position Velocity Diagrams} \label{sec:pvd}
PVDs provide more precise information on velocity and position than channel maps for a given emission component, because a channel map is produced by collapsing the datacube along velocity axis, losing detailed information otherwise attainable from a PVD.
Figure \ref{fig:pvd}$-$\ref{fig:pvd_sum} show the PVDs of the H$_{2}$ $\lambda$2.122 $\micron$ 1$-$0 S(1) and {\FeII} $\lambda$1.644 $\micron$ emission lines. In both figures white contours represent the H$_{2}$ emission and the [Fe II] emission is shown as the color intensity map. 
The slit P.A. is 225$\degree$ in all PVDs we show in this study. In \citetalias{Oh2016a}, the slit P.A. was 256$\degree$. The dash-dotted vertical lines mark the systemic velocity at $\VLSR$ $=$ $-$11.5 {\kms}.
We described the PVDs covering knot A and B in HH 167 in \citetalias{Oh2016a} in detail. The overall results for those two knots are similar to that of \citetalias{Oh2016a}.
In Figure \ref{fig:pvd}, the PVDs from slit positions `a'$-$`f' marked in Figure \ref{fig:slitposition} are shown.
In panel (b), (c) and (d), we mark the peaks B, A2$\arcmin$, and A1$\arcmin$, respectively.
Figure \ref{fig:pvd}(c) shows that the peak velocity of redshifted emission of peak A2$\arcmin$ seen in channel maps (black ellipses) is $\VLSR$ $\sim$ $-$5 {\kms} ($V$$_{\rm sys}$ $=$ $-$11.5 {\kms}), at Y$=$ $-$2$\arcsec$.
In panel (b), knot B shows peak velocity at Y$=$ $-$6$\arcsec$ with $\VLSR$ $\sim$ $-$15 {\kms}, as in \citetalias{Oh2016a}.
Knot C in H$_{2}$ shows peak velocity at $\VLSR$ $=$ $-$81 {\kms} and Y$=$ $-$10\farcs5.
At the position of knot A, {\FeII} emission peaks at $-$113 {\kms} in slit position `e'. Knot C in {\FeII} shows the highest velocity among of HH 167 with $\VLSR$ $\sim$ $-$150 {\kms}.

Figure \ref{fig:pvd_sum} shows the integrated PVD of all slit positions in the spectral mapping (positions `a'$-$`f'  in Figure \ref{fig:slitposition}).
Dotted and dashed lines indicate two different trends traced by low- and high-velocity components in H$_{2}$ emission.
Along the dotted line, the high-velocity {\FeII} emission is well traced by H$_{2}$ line in both position and velocity. In contrast, very weak {\FeII} emission is detected along the dashed line.
The gradient in radial velocity ($\delta$$v$/$\delta$$l$) in dotted and dashed lines are $\sim$ 3 {\kms} arcsec$^{-1}$ and 5 {\kms} arcsec$^{-1}$, respectively.

\subsection{Redshifted Knots to the Northeast of LkH$\alpha$ 234} \label{sec:red_counterjet}
We have obtained the first near-IR spectra of the redshifted knots to the northeast of LkH$\alpha$ 234, which are the counterparts of the blueshifted HH 167 jet. 
The location of the knots correspond to the position of redshifted CO lobe shown in \citet{Edwards1983}.
We observed three selected positions, SP1$-$3 shown in Figure \ref{fig:slitposition}. Figure \ref{fig:pvd_redshift} shows the PVDs of H$_{2}$ 1$-$0 S(1) emission lines from the three positions.
SP1 is located $\sim$ 65$\arcsec$ northeast of LkH$\alpha$ 234. The peak velocity at this position is $\sim$ $-$5 {\kms} which is close to the systemic velocity ($V_{\rm sys}$ $=$ $-$11.5 {\kms}), but $\sim$ 5 {\kms} redshifted.
As shown in Figure \ref{fig:slitposition}, the distances from LkH$\alpha$ 234 to SP2 and SP3 are 14$\arcsec$$-$16$\arcsec$ at a P.A. difference of about 20$\degree$.
The PVDs from SP2 and SP3 show that they are also different in velocity.
SP2 shows multiple peaks in velocity with the lower velocity component peaking at $\VLSR$ $\sim$ $+$12 {\kms}, and the high-velocity component shows double peaks at $+$72 {\kms} and $+$102 {\kms}. The intensity of high-velocity peaks is 5 times higher than that of low-velocity peak.
This double-peak at high velocity and separated by 20$-$30 {\kms} is very typical of an unresolved bow shock internal to the fast collimated wind (e.g., L1448, \citealt{Davis1996};  DR 21, \citealt{Smith2014}). Imaging studies of HH objects with high spatial resolution show that they consist of a chain of small bow-like knots \citep[e.g.,][]{Reipurth2002,Hartigan2001}. Considering the 1.25 kpc distance of the LkH$\alpha$ 234 system, the double-peak may represent the superposition of multiple bows.
The peak velocity at SP3 is similar to that of SP1, but shows slightly higher velocity with $\VLSR$ $\sim$ $-$2 {\kms}.

\subsection{Surrounding PDR}
In addition to the redshifted knots, we observed a part of the PDR ridge to the southwest of LkH$\alpha$ 234 (SP4). The slit position is $\sim$ 43$\arcsec$ south from LkH$\alpha$ 234, as shown in Figure \ref{fig:slitposition}. \citet{Morris2004} reported that the UV field strength at this H$_{2}$ ridge is comparable to that of the reflection nebula NGC 7023, from the polycyclic aromatic hydrocarbon band intensities using $Spitzer$ Infrared Spectrograph \citep[IRS,][]{Houck2004}.
Figure \ref{fig:pvd_pdr} shows the PVD and line profile of H$_{2}$ 1$-$0 S(1) line obtained from the PDR ridge. 
The velocity width in line profile corresponds to the spectral resolution element ($\sim$ 7 {\kms}) at systemic velocity of $-$11.5 {\kms}. The widths of all detected lines are narrow and are consistent with the typical values from a normal PDR, which is $V_{\rm FWHM}$ $<$ 5 {\kms} \citep[e.g.,][]{Hogerheijde1995}.


\subsection{H$_{2}$ Line Ratios}
\subsubsection{Extinction measurement}\label{sec:Av}
We estimate the extinction around LkH$\alpha$ 234 by employing the line ratios of H$_{2}$ lines that arise from the same upper level. In this study, the pair of $v$ $=$ 1$-$0: Q(3) $\lambda$2.424 $\micron$ / S(1) $\lambda$2.122 $\micron$ is used because these lines are not impacted by telluric absorption or OH emission. 
Using the transition probabilities of two lines taken from \citet{Turner1977}, the intrinsic intensity ratio of the Q(3)/S(1) lines is given as $\sim$ 0.7.
For the calculation of the visual extinction ($\Av$) using given predicted ($R_{\rm p}$) and observed ($R_{\rm o}$) line ratios, we adopted the equation from \citet{Petersen1996}:
\begin{equation}
\Av= \frac {2.5 \log \ (R_{\rm o}/R_{\rm p})} {A_{\rm \lambda_2}/\Av - A_{\rm \lambda_1}/\Av}  \ \rm mag
\end{equation}

In the equation above, $\lambda_1$ and $\lambda_2$ are the wavelengths of Q(3) and S(1) lines, respectively. We use the extinction law $A_{\rm \lambda}$ = $\Av$(0.55$\micron$/$\lambda$)$^{1.6}$ \citep{Rieke1985} to calculate the $A_{\rm \lambda}$/$\Av$. The uncertainties of the line fluxes are propagated from the pixel variance given by the PLP. The channel maps of the 1$-$0 Q(3) and S(1) lines are smoothed with a gaussian mask of 2 $\times$ 2 pixels to reduce the impact of noise in the calculation.

For the spectral mapping area, Figure \ref{fig:Av} shows the $\Av$ in the form of channel maps over velocity range of $-$160 $<$ $\VLSR$ $<$ $+$40 {\kms}. 
We also plot the signal-to-noise ratios (S/N) of the Q(3)/S(1) line ratio and the uncertainties in the $\Av$ measurement for each channel map in Figure \ref{fig:Av}.
In the extinction plot, we excluded pixels with S/N $<$ 3 in the calculated line ratio.
Within the plotted regions, the uncertainties in the $\Av$ measurements are in large range of 0$-$15 mag.
The $\Av$ varies largely in position and velocity in the range of 0 $<$ $\Av$ $<$ 40, which corresponds to $\Ak$ $=$ 0$-$4.6.
The extinctions in the blueshifted emission with high velocity in range of $-$160 $<$ $\VLSR$ $<$ $-$80 {\kms} are higher than that of lower velocity ($-$70 $<$ $\VLSR$ $<$ $+$40 {\kms}), showing $\Av$ higher than 35 in some regions.
We note, however, most of the high extinction pixels shown in yellow in Figure \ref{fig:Av} are on the boundary of unplotted regions and their uncertainties are relatively higher. We suspect that the calculated extinction values in these regions are less reliable.
The extinction is relatively small (0 $<$ $\Av$ $<$ 10) in low velocity range. At the left side of the figure, we also show $\Av$ in the velocity integrated image.
Similar to those in low-velocity, the extinction in the integrated map varies between 0 and 10, except for the yellow part with low reliability. This is an expected result because most of the flux in the integrated map is from low-velocity.
We also estimate the extinction for SP1$-$SP4 using the same H$_{2}$ line ratios. Extinction at the redshifted knot SP1 is highest ($\Av$ $\sim$ 21). SP2 and SP3 show smaller values ($\sim$ 7 and 16, respectively).
At SP4, the PDR at south of LkH$\alpha$ 234, the measured $\Av$ is close to zero.

\subsubsection{H$_{2}$ line properties} \label{sec:H2_ratio}

The line ratios between various H$_{2}$ emission lines allow us to probe the gas properties \citep{Smith1995,Black1987}. We measure the line ratios from 8 different positions around LkH$\alpha$ 234, including redshifted and blueshfited knots, and ambient PDR.
The H$_{2}$ excitation diagrams in Figure \ref{fig:H2_ratio} show the column density ($N(v, J)$/$g(v, J)$;  where $g(v, J)$ is the statistical weight for the vibrational and rotational levels $v$ and $J$) plotted against the excitation temperature ($E$). The different colors and shapes of symbols represent different vibrational levels and the ortho/para forms of H$_{2}$ lines, respectively.
The H$_{2}$ narrow-band image in Figure \ref{fig:slitposition} shows the slit positions from which the line ratios are measured for deriving the excitation diagrams.
Also, selected position-velocity ranges are marked by boxes on PVDs in Figures \ref{fig:pvd}, \ref{fig:pvd_redshift}, and \ref{fig:pvd_pdr}.
Figure \ref{fig:H2_ratio}(a)$-$(c) (SP 1$-$3) correspond to the redshifted knots to the northeast of HH 167 and Figure \ref{fig:H2_ratio}(d)$-$(e) correspond to the knot A and B, respectively. Figure \ref{fig:H2_ratio}(f) and \ref{fig:H2_ratio}(g) are derived from the same position of knot C, but they represent different velocity components: the high-velocity shock component ($-$100 $<$ $\VLSR$ $<$ $-$50 {\kms}) and PDR close to the systemic velocity ($-$23 $<$ $\VLSR$ $<$ $+$2 {\kms}), respectively. Figure \ref{fig:H2_ratio}(h) (SP4) is part of the PDR ridge at 43$\arcsec$ south from LkH$\alpha$ 234. The reddening is corrected with $A_{\rm \lambda}$ at each position with the values indicated in Section \ref{sec:Av}. $\Ak$ value is shown in each panel in Figure \ref{fig:H2_ratio}.

We exclude the lines affected by OH sky emission or telluric absorptions, and lines with S/N ratio below 2. 
There is an observational factor affecting the numbers of data points in the different positions. 
Fewer H$_{2}$ emission lines with $v$ $=$ 1 in (a)$-$(c) are measured than in (d)$-$(e) because the extinction is stronger in redshifted knots ($\Ak$ $=$ 1.9 $-$ 2.4) than in a blueshifted region ($\Ak$ $=$ 0.1 $-$ 0.8). In $H$-band, where the effects due to extinction are much higher, we miss the H$_{2}$ lines from higher rotational level ($J$) in $v$ $=$ 1. In (f), there are less points despite weaker extinction ($\Ak$ $\sim$ 0.1) in knot C. The reasons for this are the lower S/N in knot C than knots A and B (see also Section \ref{sec:monochromatic}), and contamination by OH and telluric lines in the high-velocity range of knot C.

The outflow regions (Figure \ref{fig:H2_ratio}(a)$-$(f)) and ambient PDR (Figure \ref{fig:H2_ratio}(g)$-$(h)) show clear differences in the shape of excitation diagrams and the transition levels of detected emission lines. We detect lines with $v$ $=$ 1$-$3 in outflow regions and $v$ $=$ 1$-$9 in the PDRs. Observed H$_{2}$ populations follow signatures of typical shocked regions and PDRs, where the shocked H$_{2}$ emission represents a thermalized population being aligned on a single line in the excitation diagram \citep[e.g.,][]{Beckwith1978}, while we expect multiple vibrational temperatures from PDRs \citep[e.g.,][]{Kaplan2017}. H$_{2}$ populations are non-thermal due to UV florescence in PDR, so we are able to distinguish it from the thermal level populations found in shocked H$_{2}$.
We suspect that the emission from outflow mostly arises from the cooling of shocked gas.
In Section \ref{sec:discussion}, we discuss more on excitation diagrams in comparison with various shock models and previous observations on Orion KL.

\section{Discussion} \label{sec:discussion}
\subsection{Multiple Outflows and Sources}

\begin{figure}
\epsscale{1.15}
\plotone{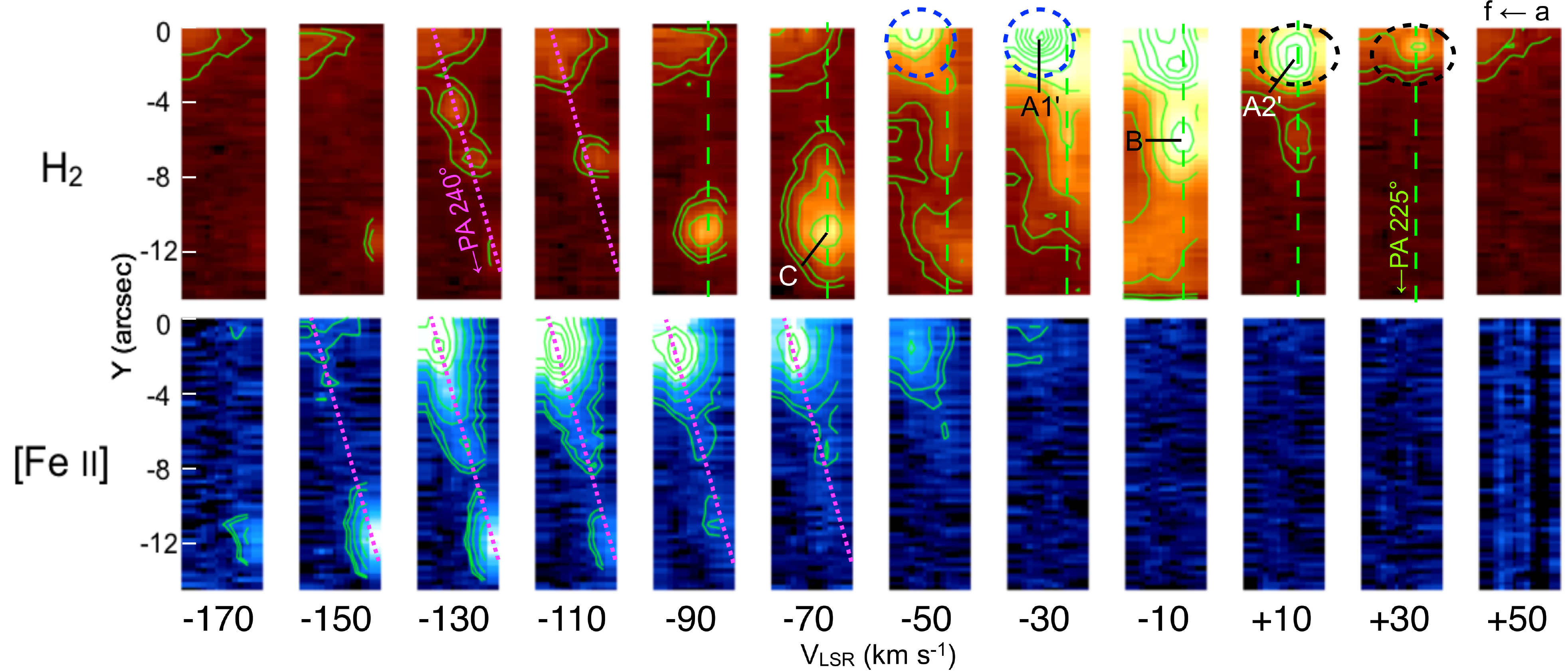}
\caption{Channel maps of the H$_{2}$ 1$-$0 S(1) $\lambda$2.122 $\micron$ line (top) and  {\FeII} $\lambda$1.644 $\mu$m (bottom). The intensity is integrated over successive 20 {\kms} intervals. The radial velocity increases from left to right, and the central velocity ($\VLSR$) in {\kms} unit is marked at the bottom of each channel map. H$_{2}$ and {\FeII} contours start from 5$\sigma$ and 3$\sigma$ levels, and increase with equal intervals in square root scale to the highest levels of 37$\sigma$ and 10$\sigma$, respectively. The slit position angle is 225$\degree$. Slit positions `a' to `f' are located from right to left in each channel map (see Figure \ref{fig:slitposition} inset at the top-right corner). The peaks `A1$\arcmin$', `A2$\arcmin$', and `B' identified in \citetalias{Oh2016a} are marked. Peak `C' corresponding to knot C is also marked. Dotted, dashed lines and circles in different colors indicate the axes and the peak positions of the identified outflows; dotted magenta lines: VLA 3B  outflow, dashed green lines: FIRS1-MM1 outflow, black ellipses: redshifted outflow from VLA 2, blue circles: an additional possible outflow from VLA 2 or 3.\label{fig:channel}}

\end{figure}

\begin{figure}
\epsscale{.5}
\plotone{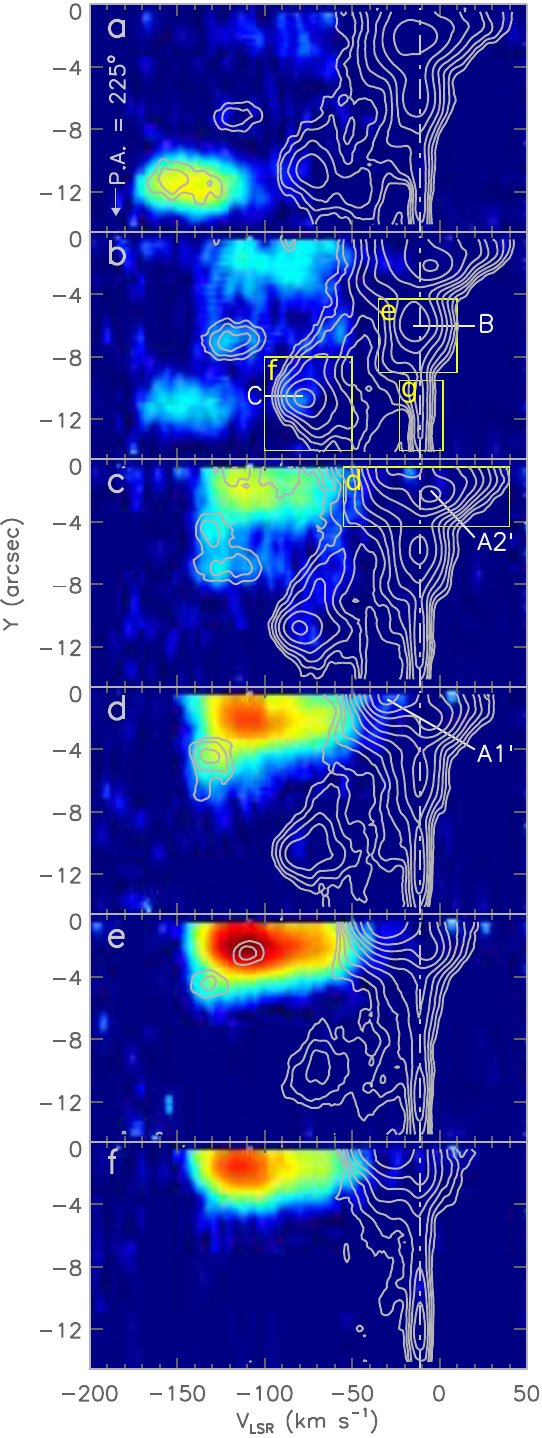}
\caption{Position-velocity diagrams of H$_{2}$ 1$-$0 S(1) $\lambda$2.122 $\micron$ (white contours) and {\FeII} $\lambda$1.644 $\mu$m (color intensity map) emission. The panels a$-$f correspond to slit positions `a'$-$`f' in Figure \ref{fig:slitposition}. The slit position anlge is 225$\degree$ in all panels, with northeast up. The contour starts from a 3$\sigma$ level, and it increases with equal intervals in a logarithmic scale. The continuum emission is subtracted. The vertical dash-dotted line indicates the systemic velocity \citep[$\VLSR$ = $-$11.5 {\kms},][]{Girart2016}. The peaks `B', and `C' are marked in (b), and `A2$\arcmin$', and `A1$\arcmin$' are marked in (c) and (d), respectively. Solid rectangles in (b) and (c) mark the regions used to calculate the excitation diagrams in Figure \ref{fig:H2_ratio}(d)$-$\ref{fig:H2_ratio}(g). \label{fig:pvd}}
\end{figure}

\begin{figure}
\epsscale{.5}
\plotone{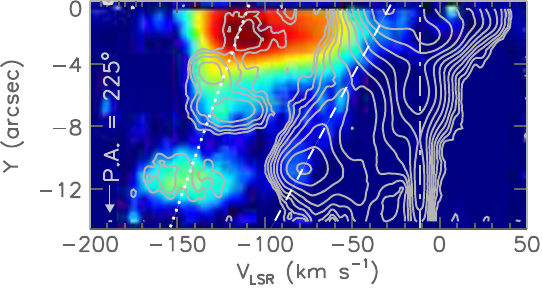}
\caption{Integrated position-velocity diagrams of H$_{2}$ 1$-$0 S(1) $\lambda$2.122 $\micron$ (white contours) and {\FeII} $\lambda$1.644 $\mu$m (color intensity map) emission. Northeast is up and southwest is down in the diagrams (P.A. $=$ 225$\degree$). The PVDs at slit positions `a'$-$`f' are summed to show overall kinematic information in the spectral mapping area. The contour starts from a 3$\sigma$ level, and it increases with equal intervals in a logarithmic scale. The background continuum emission is subtracted. Dotted and dashed lines indicate the velocity gradients of different outflows associated with VLA 3B and FIRS1-MM1, respectively.
The vertical dash-dotted line marks systemic velocity ($\VLSR$ = $-$11.5 {\kms}). \label{fig:pvd_sum}}
\end{figure}

\begin{figure}
\epsscale{.5}
\plotone{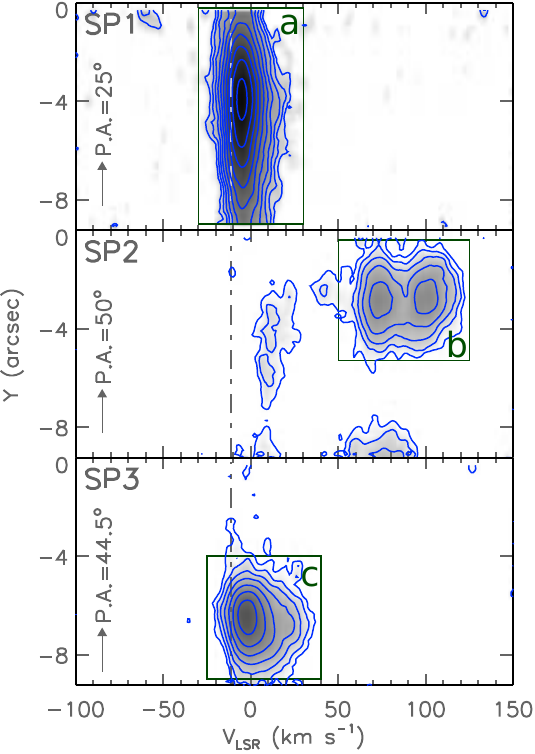}
\caption{PVDs of H$_{2}$ 1$-$0 S(1) $\lambda$2.122 $\micron$ of redshifted knots at northeast of LkH$\alpha$ 234. Top, middle, and bottom panels correspond to the SP1, SP2, and SP3 in Figure \ref{fig:slitposition} with the slit P.A. of 25$\degree$, 50$\degree$, and 44\fdg5, respectively.  The contour starts from a 3$\sigma$ level, and it increases with equal intervals in a logarithmic scale. Solid rectangles mark the regions used to derive the excitation diagrams in Figure \ref{fig:H2_ratio}(a)$-$\ref{fig:H2_ratio}(c). \label{fig:pvd_redshift}}
\end{figure}

\begin{figure}
\epsscale{.5}
\plotone{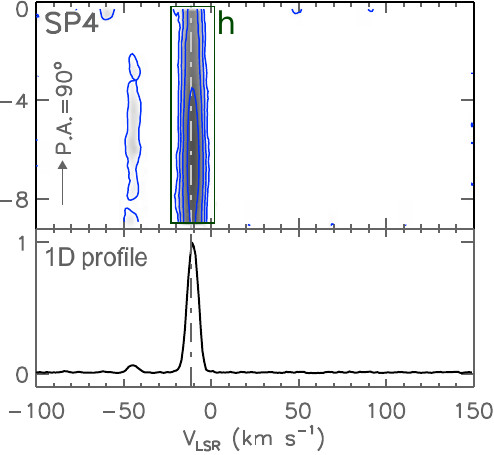}
\caption{PVD (top) and line profile (bottom) of H$_{2}$ 1$-$0 S(1) $\lambda$2.122 $\micron$ obtained at SP4 (``PDR ridge''). The slit position is shown in Figure \ref{fig:slitposition}. V$_{FWHM}$ is  $\sim$ 7 {\kms} at systemic velocity, which corresponds to the resolution limit of the instrument. The weak component detected at V$_{LSR}$ $=$ $-$45 {\kms} is emission from H$_{2}$ 8$-$6 O(4) transition, also arises in the PDR. The region within the solid rectangle is used to calculate an excitation diagram in Figure \ref{fig:H2_ratio}(h). East is up and south is down in the diagram. \label{fig:pvd_pdr}}
\end{figure}

\begin{figure}
\epsscale{1.15}
\plotone{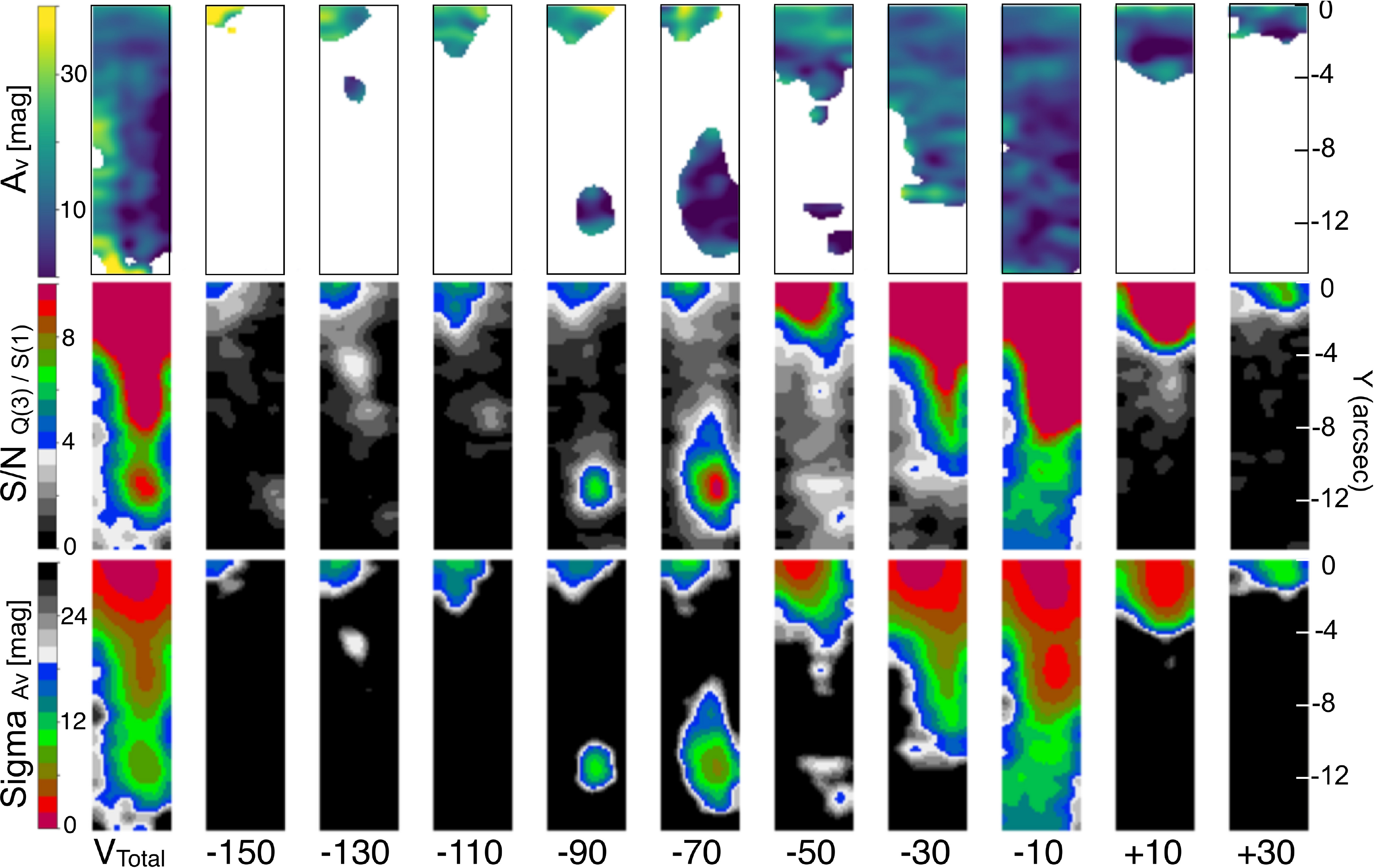}
\caption{Visual extinctions ($\Av$) of channel maps derived from a line ratio of H$_{2}$ $v$ $=$ 1$-$0: Q(3) / S(1) (top), the signal-to-noise ratio (S/N) of the line ratio (middle), and the uncertainty in the $\Av$ measurement (bottom). The P.A. is 225$\degree$, with northeast at the top. The central velocity is marked at the bottom of each channel map. The panels at the most left side show the maps with integrated velocity in $-$160 {\kms} $<$ $\VLSR$ $<$ $+$40 {\kms}. Pixels with S/N $<$3 in the line ratio are excluded in the $\Av$ plot. The channel maps of Q(3) and S(1) lines are smoothed with a gaussian mask of 2 $\times$ 2 pixels before the calculation of the extinction.
\label{fig:Av}}
\end{figure}

\begin{figure}
\epsscale{.9}
\plotone{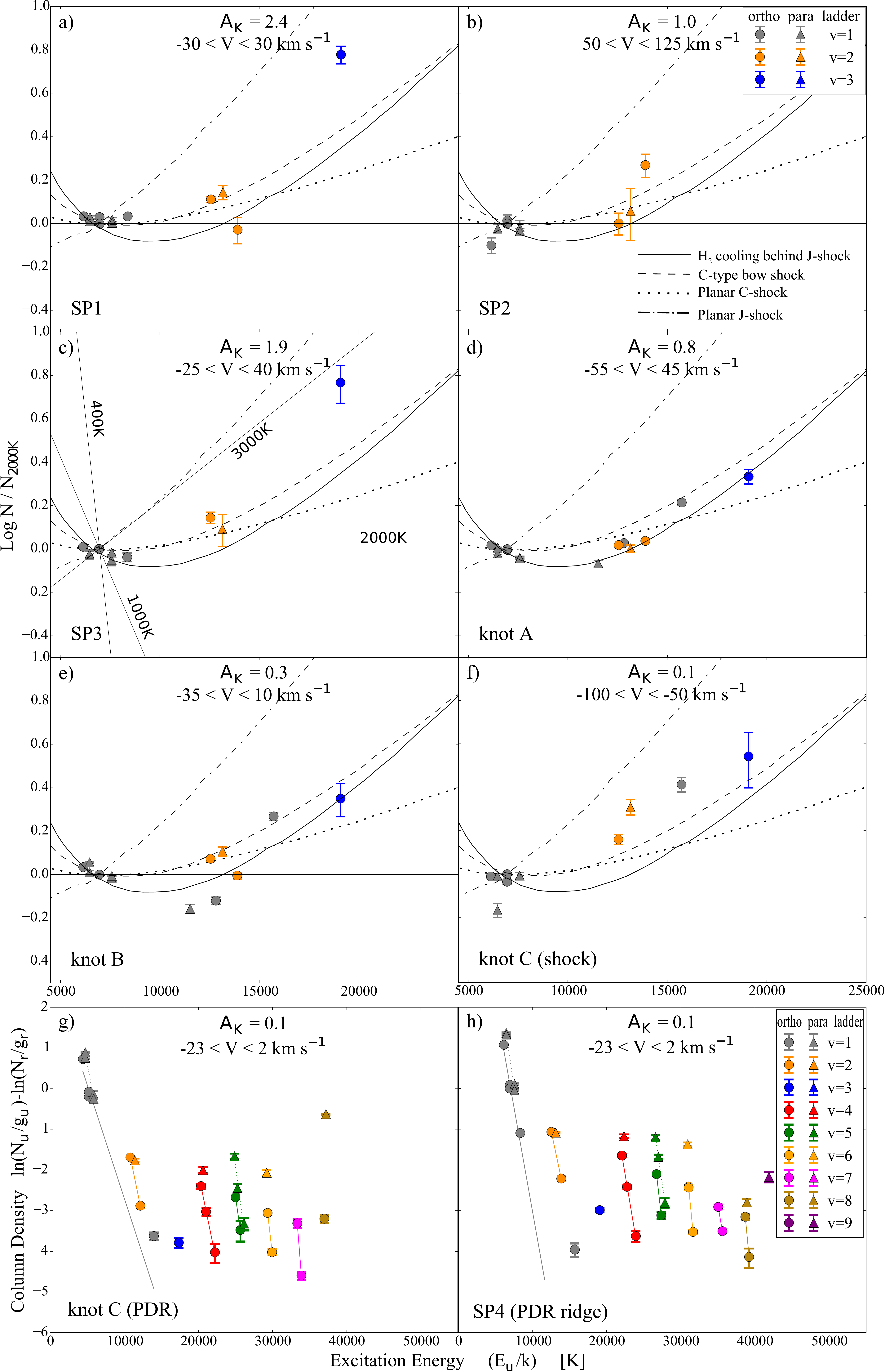}
\caption{\scriptsize{H$_{2}$ excitation diagrams at 8 different positions around the LkH$\alpha$ 234 outflow. Diagrams are from (a)$-$(c) redshifted knots (SP 1$-$3), (d) knot A, (e) knot B, (f) high-velocity shock component and (g) PDR at systemic velocity in knot C of HH 167, and (h) PDR ridge (SP4). The slit positions are shown in the H$_{2}$ narrow-band image (Figure \ref{fig:slitposition}) and their position-velocity ranges are marked in Figures \ref{fig:pvd}, \ref{fig:pvd_redshift}, and \ref{fig:pvd_pdr}. The extinction ($\Ak$) and velocity range are shown in each panel. In all panels, the column densities are normalized to H$_{2}$ 1$-$0 S(1) line. In (a)$-$(f), plots are relative to the Boltzmann distribution at 2000 K. The model plots are indicated$-$ Solid: H$_{2}$ cooling zone after $J$-shock \citep{Brand1988,Burton1989}, dashed: $C$-type planar shock \citep{Smith1991}, dotted: $C$-type bow shock model \citep{Smith1991}, and dash-dotted: planar $J$-shock model with conventional cooling \citep{Smith1991,Burton1989}. In (c), populations at single temperatures of 400, 1000, 2000 and 3000 K are shown with solid straight lines.} \label{fig:H2_ratio}}
\end{figure}

In Section \ref{sec:channel}, we showed four different emission features traced by lines and circles in Figure \ref{fig:channel}.
The origin of these emission features can be interpreted as either a single outflow, even if they are not aligned perfectly with each other, or multiple outflows originating from multiple sources.
As a basis for the single outflow case, there are examples of flows with kinks and curves in them and poorly collimated jets (e.g., HH jets in NGC 1333, \citealt{Bally2001}; HL/XZ tau jet, \citealt{Movsessian2007}; and L1660, \citealt{Davis1997}). It is often the case that a single outflow produces two very different phenomena- an internal shock in the fast jet that shows up at high velocity emission and a shock of the swept-up material into the ambient cloud that appears close to the ambient cloud velocity \citep{Pyo2003,Bally2007}. Also, positions and velocities of H$_{2}$ and {\FeII} emission could be different when they arise from a single outflow since they originate from different shock regions \citep{Pyo2002,Davis2003}.

However, we prefer the interpretation with multiple outflows through the identification of at least three separate outflows. 
The channel maps in Figure \ref{fig:channel} show that different kinematic features have different velocity-space vectors in a 3D geometric space, and there are multiple YSO candidates within 5$\arcsec$ of the bases of the different outflow vectors.
In addition, sub-millimeter and radio observations have reported on multiple outflows in this area \citep{Trinidad2004,Girart2016}.
We argue that the multiple outflows originate from VLA 3B, MM1, and VLA 2. Below we describe the axes and driving source of each outflow with their kinematic properties found in the channel maps and PVDs.

\begin{figure}
\epsscale{.9}
\plotone{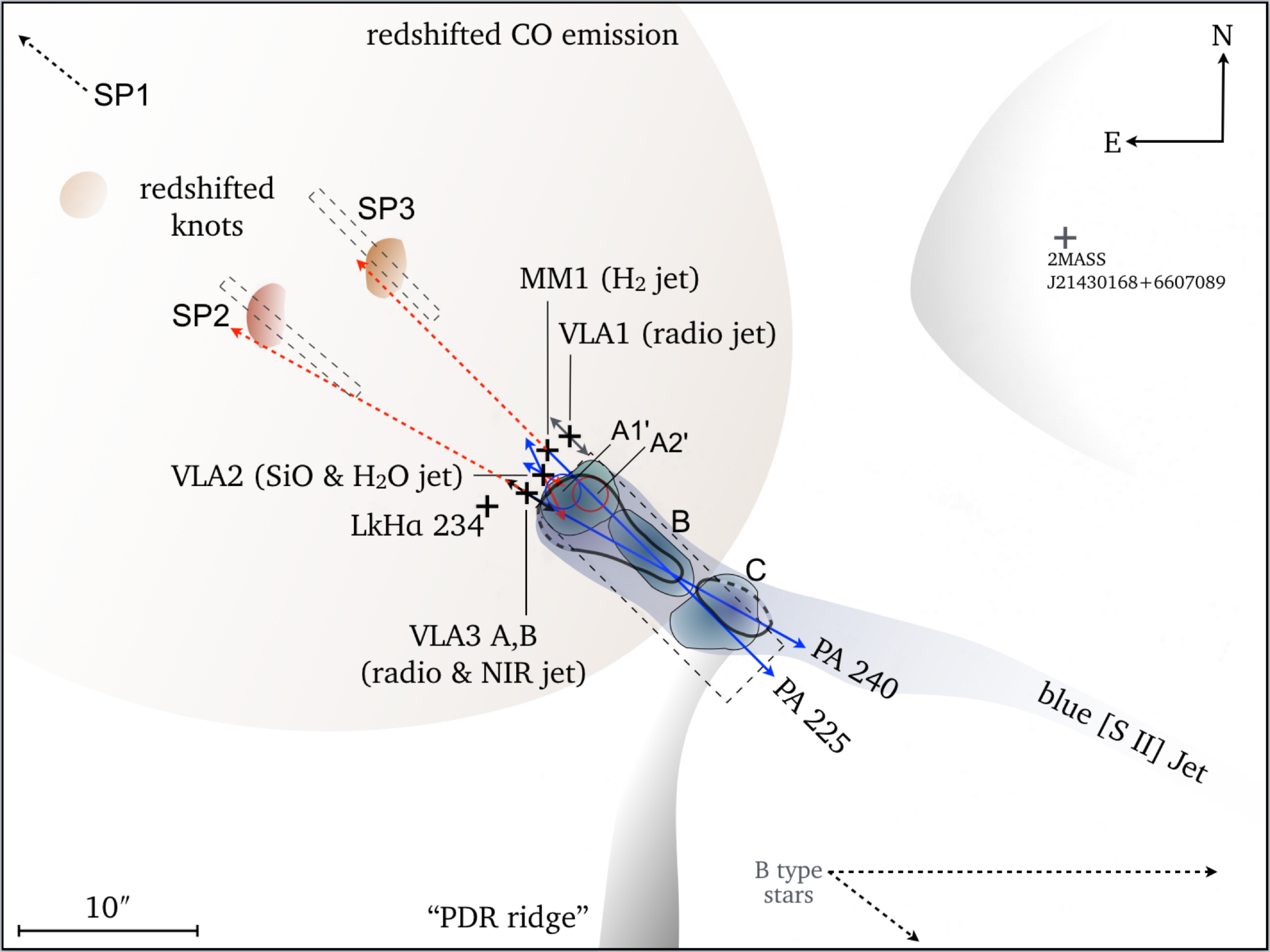}
\caption{Schematic diagram of multiple outflows around LkH$\alpha$ 234. The crosses and solid arrows show the locations of YSOs and axes of the outflows, respectively. The regions within thin and thick solid lines represent 3$\sigma$ level of H$_{2}$ 1$-$0 S(1) and {\FeII} $\lambda$1.644 $\micron$ emission, respectively. The peaks A1$\arcmin$, A2$\arcmin$ and knots B, C are marked. The area of spectral mapping is marked with a dashed rectangle. The position of blueshifted {\SII} jet is taken from \citet{Ray1990}. Two red, dashed arrows are extensions of the blueshifted emission, which has PA of 240 and 225. \label{fig:schematic}}
\end{figure}

In Figure \ref{fig:schematic}, we present a schematic drawing showing a overall distribution of outflows and YSO candidates around LkH$\alpha$ 234. The outflow axes and positions of the emission peaks are also indicated in the figure.
Figure \ref{fig:channel} showed that the axis with P.A. $=$ 240$\degree$ traces strong {\FeII} emission and weak H$_{2}$ emission. It is aligned well with the position of radio continuum source VLA 3B, indicating that this outflow is driven by VLA 3B.
This result agrees with the idea from the previous study in \citetalias{Oh2016a}, which argued that the high-velocity {\FeII} jet ($\VLSR$ $=$ $-$120 {\kms}) is driven by radio continuum source VLA 3B, based on its positional agreement with the radio thermal jet around VLA 3B \citep{Trinidad2004} which has P.A. of $\sim$ 230$\degree$. The inclination angle of this outflow is unknown. We assume that this outflow is toward us rather than close to the sky plane, because the observed radial velocity is highest with $\sim$ $-$150 {\kms} among the sources in this area.

The axis with P.A. of 225$\degree$ traces the H$_{2}$ jet, which has been observed in the narrow-band H$_{2}$ image by \citet{Cabrit1997}. The axis of this jet is aligned well with the embedded source FIRS1-MM1 \citep{Fuente2001,Girart2016}.
This outflow is traced in all three knots. The blueshifted peaks B and C, and the peak A2$\arcmin$ are in good agreement with this axis. We note that A2$\arcmin$ is redshifted with respect to the systemic velocity ($V_{\rm sys}$ $=$ $-$11.5 {\kms}).
In \citetalias{Oh2016a}, we noted this redshifted component with low-velocity ($\VLSR$ $<$ $+$50 {\kms}) at the position of knot A and we proposed that it is an outflow with a wide-opening angle with the axis close to the sky plane.
With an assumption that this emission comes from a part of FIRS1-MM1 jet, it could arise from a wide-angle outflow showing both blue and redshifted components close to the source.
On the other hand, we argue a possible association between this redshifted H$_{2}$ emission and the radio continuum source VLA 2.
H$_{2}$O masers have been detected around VLA 2 \citep{Trinidad2004,Marvel2005,Torrelles2014}. The proper motion and the locations of the masers indicate that VLA 2 has a bipolar outflow in the northeast-southwest direction, with the redshift lobe toward southwest. \citet{Torrelles2014} estimated the inclination angle of the outflow as about 15$\degree$, with respect to the plane of the sky. \citet{Girart2016} also reported the larger scale SiO outflow which might be a counterpart of the H$_{2}$O masers. Peak A2$\arcmin$ is located on the extension of the axis of H$_{2}$O masers, which is aligned with FIRS1-MM1 jet, too.

From the discussion above, we suggest several possible origins of the emission from A2$\arcmin$. Firstly, the emission could originate from the outflow driven by VLA 2. In this case, the emission is a newly found near-IR counterpart of that outflow.
Secondly, it could be part of the FIRS1-MM1 jet. Thirdly, the multi-contribution of both VLA 2 and FIRS1-MM1 could cause the emission in knot A. Peak A2$\arcmin$ could be a place where two outflows with different axes are interacting each other, with the bright shock emission could arise.

The driving source of the H$_{2}$ emission in peak A1$\arcmin$ is not clear with given positional information. We assume it is originated from VLA 2 or 3 \citep{Trinidad2004,Girart2016}, based on its proximity to those sources. An additional study with higher spatial resolution would help further our understanding of the origin of this shocked emission.

Optical {\SII} emission traces axes of both 240$\degree$ and 225$\degree$ \citep{Ray1990}.
In \citetalias{Oh2016a}, we assumed that the optical {\SII} jet is driven by VLA 2. The origin of {\SII} outflow, however, is not clear because it shows different axis in inner and outer regions \citep{Ray1990}. High-resolution narrow-band imaging around the central sources is required to reveal their origin. We should not rule out the possibility that {\SII} emission arises from multiple outflows, and the source of emission in inner and outer regions could be different.
On the other hand, we show that the line ratios for H$_{2}$ indicate largely variable $\Av$, in the range of 0 $<$ $\Av$ $<$ 40 in the blueshifted components, although the uncertainties are larger in the regions with higher $\Av$. The presence of optical lines from shocks in this region is a clear indication that extinction is highly variable, perhaps also clumpy.

The blueshifted emission in the high-velocity of $\VLSR$ $>$ $-$100 {\kms} with a large variation in $\Av$ are associated with VLA 3B. 
In contrast to the high-velocity components, extinction and its variation are relatively small (0 $<$ $\Av$ $<$ 10) in velocity range of $-$70 $<$ $\VLSR$ $<$ $-$10 {\kms}
where the emission is mostly from the MM1 outflow (see Figure \ref{fig:channel} and \ref{fig:Av}). The redshifted outflow which is probably driven by VLA 2 also shows a small extinction value, with $\Av$ $<$ 10.
This extinction trend contrasts with the Orion KL outflow discussed in \citet{Oh2016b}, which showed relatively small extinction ($\Av$ $\sim$ 0) at the highest blueshifted velocity and larger $\Av$ values at redshifted velocities, which implied a differential extinction along the line of sight.
Orion KL is a huge outflow system with more than a hundred bullets ejected radially from a single origin \citep{Bally2015,Youngblood2016}.
The LkH$\alpha$ 234 system, however, likely shows a more complex distribution of extinction because different sources emanate multiple outflows with different axes.
In addition, the distributions of dust filaments detected by \citet{Girart2016} showed that the dust emission is strongest around VLA 3 and is weaker in the northwest. The extinction in channel maps showed that $\Av$ is highest in the southeast (the upper left of the channel maps), which is in agreement with the dust distribution shown in \citet{Girart2016}.

\subsection{Multiple Counter Jets}
In Section \ref{sec:red_counterjet}, the redshifted knots observed in SP1$-$3 show different velocities. The emission from SP2 and SP3 show an especially large difference in the peak velocity, $\VLSR$ $\sim$ $+$100 and $-$2 {\kms}, despite their similar distance from the sources.
This is an additional argument for multiple outflows indicating that they are the red counterparts of the multiple blueshifted outflows, because the outflows usually are bipolar phenomena.
From from its similarity in the velocity found in this study, the high-velocity knot in SP2 with $\VLSR$ $\sim$ $+$100 {\kms} is probably the redshifted counterpart of the blueshifted jet driven by VLA 3B with $\VLSR$ $\sim$ $-$100 {\kms}. The position of this knot, which is at P.A. $\sim$ 35$\degree$ from VLA 3 (see Figure \ref{fig:schematic}), also supports that this could be a counterpart of the VLA 3B jet. In the case of the knot at SP3, it can be matched with two counter outflows, the MM1-jet and the radio jet of VLA 1 \citep{Trinidad2004}. Since the SP3 knot is located at P.A. $=$ 40$\degree$ from MM1, we suppose it is most convincing that this is a red counterpart of the MM1-jet with P.A. $=$ 225$\degree$. Its low radial velocity ($\VLSR$ $\sim$ $-$2 {\kms}) also support this idea, given that the MM1-jet is dominant in low-velocity. SP3 knot is also on the extended axis of radio jet detected around VLA 1 with P.A. $=$ 45$\degree$ \citep{Trinidad2004}, indicating that VLA 1 could be another candidate for the origin of this redshifted emission.

The driving source of the knot at SP1 could be the same as of SP2, because they show similar velocities. However, SP1 is located much farther from the YSO candidates than SP2. We note that the shocked H$_{2}$ emission at SP1 is spread over very wide region, as shown by \citet{Eisloffel2000}.

%

\subsection{Shocked H$_{2}$ Gas}

\subsubsection{Excitation diagram}
From the excitation diagrams in Figure \ref{fig:H2_ratio}, we showed that the emission from outflow regions arises from the cooling by shocked H$_{2}$ gas.
The model calculations of various shocks are overplotted on Figure \ref{fig:H2_ratio}(a)$-$(f). Solid line indicates an empirical model with planar $J$-shock cooling flow assuming that H$_{2}$ lines dominate the cooling \citep{Brand1988,Burton1989}. Dashed and dotted lines are $C$-type planar shock and bow-shock models \citep{Smith1991}, respectively. For the bow-shock model, we assumed the bow speed is 180 {\kms}, n$_{\rm H}$ = 3 $\times$ 10$^{6}$ cm$^{-3}$, and an ionization fraction is 3 $\times$ 10$^{-7}$ \citep{Smith1991}. Dash-dotted line shows a planar $J$-type shock model with conventional cooling \citep{Smith1991,Burton1989}.
A single model does not well-reproduce the populations except for knot A (Figure \ref{fig:H2_ratio}(d)), which  shows reasonable agreement with the $J$-shock cooling flow model \citep{Brand1988,Burton1989}.
We note that discrimination between different models is not feasible in the excitation energy range below 10000 K.

The population of knot B is scattered in Figure \ref{fig:H2_ratio}(e) toward the bottom of the figure at excitation energy over 10000 K. This scatter is not shown in knot A (Figure \ref{fig:H2_ratio}(d)). This scatter might be caused by the mixture of emission from both outflow and PDR, because the line intensity is integrated from low-velocity ($-$35 $<$ $\VLSR$ $<$ $+$10 {\kms}).
The population in the high-excitation range of knot C ($>$10000 K) show a trend toward planar $J$-shock compared to other panels, indicating a possible shock-type mixture. {\FeII} emission with higher-velocity detected at the position of knot C (see Figure \ref{fig:pvd}(b)) supports this idea, which usually arises from $J$-shock regions \citep[e.g.,][]{Pyo2002,Koo2016}.
In the case of the Orion KL region, associated with the formation of massive stars, the shock-type mixture is also suggested. The H$_{2}$ population from Orion KL follows two models showing similar populations: H$_{2}$ cooling flow behind $J$-shock and bow-shock models \citep{Oh2016b}. The smaller number of data points in this study limits more detailed comparison in line ratios between around LkH$\alpha$ 234 and the Orion KL.

The analysis with the H$_{2}$ excitation diagram shows that only with spectroscopy can we inventory the amount of shocked material in a definitive way.
We show that some of the emission seen in the imaging study is very noticeably PDR. In the case of knot C, even in the same location in image, two different, velocity-resolved components represent shocks and PDR, respectively (see Figure \ref{fig:H2_ratio}(f)$-$(g)). There have been several big H$_{2}$ surveys; \citet{Froebrich2011} from UKIRT, and many others, for example, and \citet{Yu1997}, \citet{Eisloffel2000}, \citet{Walawender2013}. This illustration points out that spectroscopy is a critical tool to confirm whether the features identified in imaging surveys represent one or the other of these two very different physical phenomena.

\subsubsection{High-velocity H${_2}$ emission}

The maximum radial velocity of H${_2}$ gas in HH 167 reaches over $-$150 {\kms} (see Figures \ref{fig:channel}$-$\ref{fig:pvd_sum}). This velocity is much greater than the critical shock speed (40$-$50 {\kms}) at which H${_2}$ molecules are dissociated \citep{Draine1993,Hollenbach1997}. Such ``supercritical-velocity'' (SCV hereafter) H${_2}$ outflows have been observed in many sources (e.g., L1551 IRS5, \citealt{Davis2003}; HH7, \citealt{Pike2016}; Orion KL, \citealt{Oh2016b}; DR 21, \citealt{Smith2014}), and several explanations have been proposed.
In the following, we briefly summarize these explanations and then discuss the origin of the SCV H${_2}$ outflow in HH 167.

The proposed explanations for the origin of the SCV H${_2}$ outflow may be divided into two categories depending on whether the H${_2}$ is assumed to be in the jet or in the swept-up ambient medium.\footnote{The term `outflow' often includes both the outflow/jet/wind originated from the young stellar object and the swept-up ambient medium. Here, in order to avoid a confusion, we will use the term `jet' for the outflow/jet/wind originated from the young stellar object.} In the former explanation, the jet is molecular and the emission is from H${_2}$ molecules in the jet heated by slow internal shock by ``self-shocking''. That is, a faster wind component ``catches up'' to a portion of the wind that either was slower to begin with or has slowed down.
Behind the bow shock, the emission can also be driven by the reverse shock due to the interaction with ambient clouds \citep{Davis2002,Smith2014}.
In the latter explanation, the emission is from H${_2}$ molecules in the swept-up ambient medium. They could be either pre-existing molecules that are not dissociated or new molecules formed after the dissociation. The ambient molecular gas can be accelerated to SCV ($\ge$50 {\kms}) without dissociation if the Alfv\'{e}n speed of the medium is very large \citep{Smith1991} or if the acceleration occurs gradually by a slowly accelerating jet \citep{Lim2002}. On the other hand, molecules can reform in the postshock cooling layer in fast dissociative shock \citep{Hollenbach1989}. In the latter case, since the reformation proceeds at temperatures $\lesssim$ 500 K, the H${_2}$ 2.122 $\micron$ emission line would be weak and the H${_2}$ spectra is characterized by the formation pumping of H${_2}$ so that the line ratio 1$-$0 S(1)/2$-$1 S(1) should be an order of $\sim$ 2$-$3 \citep[][see also \citealt{Pike2016} and references therein]{Hollenbach1989}.

\begin{figure}
\epsscale{.5}
\plotone{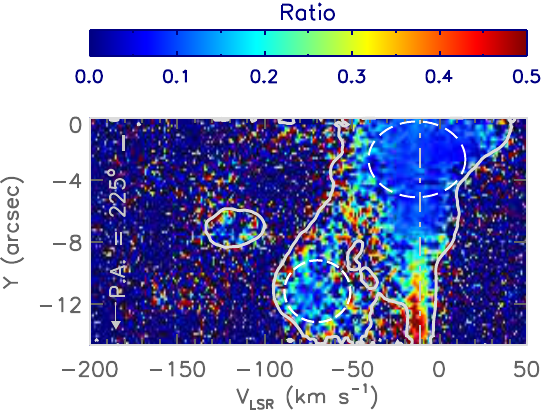}
\caption{PVD of the line ratio of H$_{2}$ 2$-$1 S(1) / 1$-$0 S(1) at the selected slit position `b' from the spectral mapping of HH 167. The solid lines indicate a 3$\sigma$ level of 1$-$0 S(1) line. The regions within the dashed circles are used for the measurement of line ratios of low- and high-velocity components in Figure \ref{fig:excitation_diagram}. In the case of the pure $C$-shock, the ratio is close to 0.05 while for the $J$-shock it becomes $\sim$0.24, according to the model calculations of \citet{Smith1995}. \label{fig:ratio_pvd}}
\end{figure}

\begin{figure}
\epsscale{1.2}
\plotone{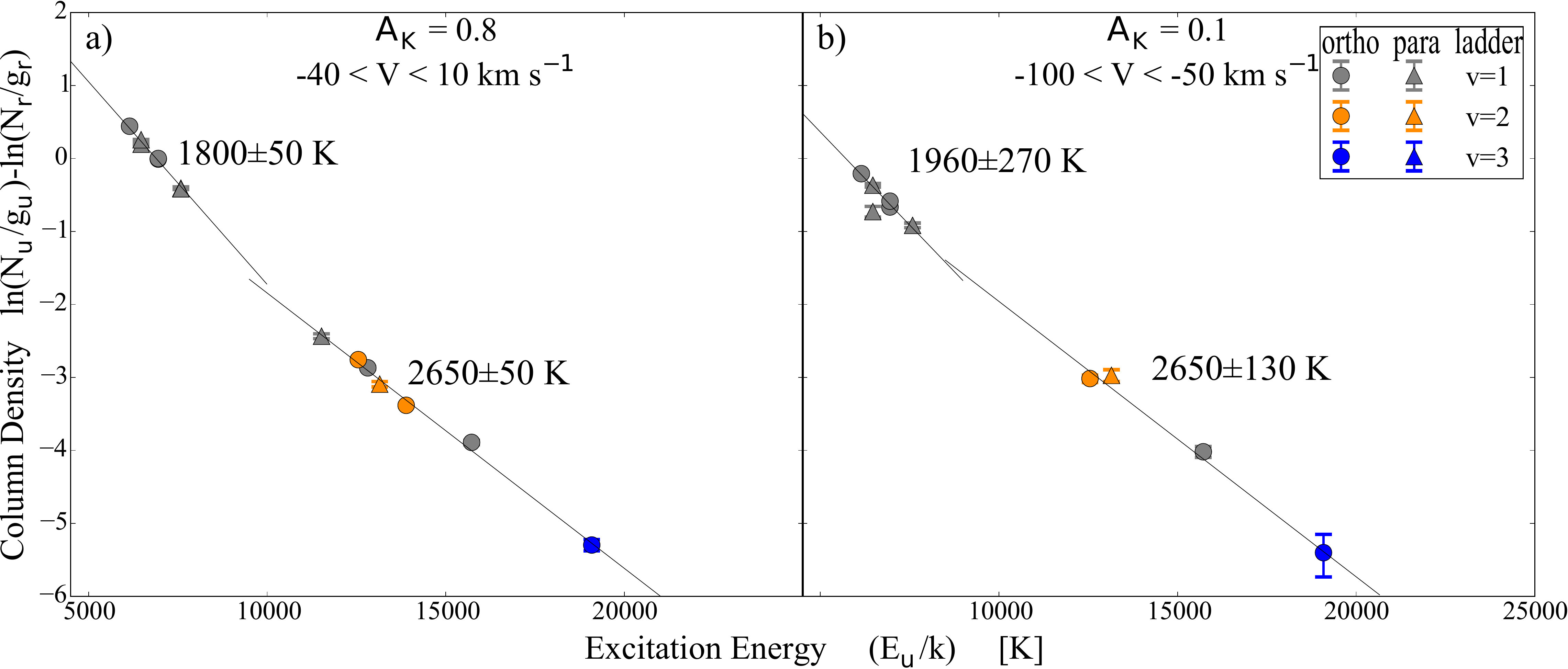}
\caption{H$_{2}$ excitation diagrams from (a) $low-$ and (b) $high-$velocity components in slit position `b' shown in Figure \ref{fig:ratio_pvd}. The relative H$_{2}$ intensities are measured at regions marked with the circles in Figure \ref{fig:ratio_pvd}. The intensity is integrated in the velocity range of $-$40 $<$ $\VLSR$ $<$ $+$10 {\kms} and $-$100 $<$ $\VLSR$ $<$ $-$50 {\kms}. In both panels, the populations are fitted by two temperatures. Calculated excitation temperatures ($T_{\rm e}$) are similar in (a) and (b), with $T_{\rm e}$ $=$ 1900$-$2000 K in the low excitation energy of $E_{\rm u}$ $<$ 10000 K, and $T_{\rm e}$ $\sim$ 2600 K in 10000 $<$ $E_{\rm u}$ $<$ 20000 K. \label{fig:excitation_diagram}}
\end{figure}

Figure \ref{fig:ratio_pvd} is a PVD of HH 167 showing the H${_2}$ 2$-$1 S(1)/1$-$0 S(1) ratio in slit position `b' obtained from our deep observation (see Section \ref{sec:obs_mapping}; see also Figure \ref{fig:slitposition}).
This area contains both low- and high-velocity components, and we derive the H${_2}$ 2$-$1 S(1)/1$-$0 S(1) ratio of 0.09$\pm$0.04 and 0.13$\pm$0.05 for the gas at $\VLSR$ $\sim$ 0 {\kms} and $-$80 {\kms}, respectively. For the gas at  $-$113 {\kms} we derive an upper limit of 0.16.
The ratio of the SCV gas is somewhat higher than that of the low-velocity gas, but it is much less than what we would expect for the ratio from the reformed molecules.
This can be also seen in Figure \ref{fig:excitation_diagram} which shows the excitation diagrams for the gas at $\VLSR$ $\sim$ 0 {\kms} and $-$80 {\kms}. The line ratios can be fit by two-temperature components at 1800$-$2700 K for both 0 {\kms} and $-$80 {\kms} components. So again their excitation temperatures are comparable, and there is no indication of a high excitation level population of SCV gas due to formation pumping. For comparison, \citet{Pike2016} detected a 5000 K component in HH7$-$11 and attributed it to the reformed H${_2}$ molecules. Therefore, with given data in this study, the SCV H${_2}$ emission in HH 167 is not likely from the reformed H${_2}$ molecules behind a fast $J$-shock. 

The above discussion leads us to conclude that the SCV H${_2}$ emission might be from pre-existing molecules swept-up by a non-dissociative shock either in the jet or in the ambient medium. Further observations with higher S/N ratio including H${_2}$ lines from higher excitation energy ($E_{\rm u}/K$ $>$ 30000 K) would help deeper understanding on the true origin of the observed SCV gas.

\subsection{Role of multiple outflow}
There is a literature going back many decades about outflows contributing to the destruction of cloud cores. 
\citet{Norman1980} first investigated the importance of energy injection by stellar outflows; they analyzed winds from T Tauri Stars because bipolar outflows from protostars were unknown at that time.
In addition to \citet{Norman1980}, \citet{Franco1983} and \citet{McKee1989} found that the energy injection rate by outflow was sufficient to support star-forming clouds against collapse.
Also, \citet{Offner2017} simulated that outflows drive turbulence in the core even the initial magnetic fields is strong, and indicated that the outflow entrains about three times the mass of real launched gas.
In this study and the \citetalias{Oh2016a}, we showed a cluster of low- to intermediate-mass stars, where each star has its own outflow, except LkH$\alpha$ 234 which could be in a later evolutionary stage. 
Intermediate-mass stars form in clusters where the total amount of material is fairly substantial. Unlike O-type stars, there is no ionizing radiation to drive the material away and any supernova that we might get will come after quite some time. We demonstrate, in the earliest phases, all of these stars in the cluster are working together to inject dynamical energy into the core.
It is unlikely that outflow alone is sufficient to play this role, as may not supply the random motions observed on large scales, as pointed out by \citet{Walawender2005} for example. 

\section{Summary} \label{sec:summary}

We present the results from high-resolution near-IR spectral mapping toward the multiple outflow around LkH$\alpha$ 234 star formation region. We summarize the main results as followings:

\begin{enumerate}
\item The channel maps made from a spectral mapping with high resolving power ($\Delta v$ = 7 {\kms}) provided ultra narrow-band images which allow us to distinguish multiple outflows overlapped spatially. We found that there are at least three different near-IR outflows  showing similar P.A. around LkH$\alpha$ 234, probably driven by embedded sources MM1, VLA 2, and VLA 3.
\item We showed that the H${_2}$ emission arises from complex multiple outflow. All of the outflows around LkH$\alpha$ 234 region show H${_2}$ emission, while some parts of them have counterparts in {\FeII} and/or {\SII} emission.
\item Low-velocity ($\VLSR$ $<$ $+$50 {\kms}), redshifted H${_2}$ emission at the base of HH 167 is aligned well with the H$_{2}$O masers around VLA 2. We suggest that this redshifted emission indicates an outflow from VLA 2, which may have the angle close to the plane of the sky.
\item We reconfirmed the P.A. of the {\FeII} jet previously detected in \citet{Oh2016b} as 240$\degree$, and conclude that it is driven by the radio continuum source VLA 3B. This outflow is also detected with weak H${_2}$ emission.
\item The first spectra of redshifted knots at the northeast of HH 167 were obtained. Different knots show different velocities with V$_{LSR}$ from $\sim$ 0 to $+$100 {\kms}, indicating that they are likely the counterparts of the multiple outflows.
\item We probed the origin of the high-velocity (50$-$120 {\kms}) H${_2}$ gas beyond the breakdown velocity. The faster H${_2}$ gas shows higher line ratios than slow gases implying that the shock properties of two velocity components are different, but we could not find evidence to show if the H${_2}$ arises from reformed gas. The H${_2}$ emission originating from the jet itself is still considerable.
\item Spectroscopy using ratios between many H$_{2}$ lines showed that some of the emission seen in the imaging study is clearly PDR, instead of shocked gas. This result indicates that spectroscopy is an unique tool to measure the amount of shocked material in a definitive way.

\end{enumerate}

\acknowledgments

This work used the Immersion Grating Infrared Spectrometer (IGRINS) that was developed under a collaboration between the University of Texas at Austin and the Korea Astronomy and Space Science Institute (KASI) with the financial support of the US National Science Foundation under grant AST-1229522, of the University of Texas at Austin, and of the Korean GMT Project of KASI.
This paper includes data taken at the McDonald Observatory of The University of Texas at Austin.
These results made use of the Discovery Channel Telescope at Lowell Observatory. Lowell is a private, non-profit institution dedicated to astrophysical research and public appreciation of astronomy and operates the DCT in partnership with Boston University, the University of Maryland, the University of Toledo, Northern Arizona University and Yale University.
This work also used the data from the Canada-France-Hawaii Telescope (CFHT) archive.
This work was supported by the National Research
Foundation of Korea (NRF) grant funded by the Korea
Government (MSIP) (No. 2012R1A4A1028713).

\software{Plotspec \citep{Kaplan2017}, IGRINS Pipeline Package \citep{Lee2017}, SAOImage DS9 \citep{Smithsonian2000}, IRAF \citep{Tody1986,Tody1993}}

\clearpage

\clearpage

\clearpage

\clearpage

\clearpage

\clearpage

\clearpage

\clearpage

\clearpage

\clearpage

\clearpage

\begin{deluxetable}{cccccc}
\tabletypesize{\scriptsize}
\tablecaption{Summary of Observations.\label{tbl:obs_log}}
\tablewidth{0pt}
\tablehead{
Object                  & Date         & Telescope/         & P.A.        & Exposure time & Slit size  \\
                        & (UT)         & Instrument         & ($\degree$) & (sec)         &                                  
}
\startdata
HH 167 mapping\tablenotemark{a} & 2015 Aug. 6  & HJST\tablenotemark{b} / IGRINS & 225       & 600\tablenotemark{c}           & 1\farcs0 $\times$ 15\farcs0 \\
HH 167 position `b'\tablenotemark{d} & 2017 Jun. 9  & HJST / IGRINS & 225       & 2400           & 1\farcs0 $\times$ 15\farcs0       \\
SP1                     & 2016 Nov. 19 & DCT\tablenotemark{e} / IGRINS       & 25          & 300           & 0\farcs63 $\times$ 9\farcs3            \\
SP2                     & 2016 Nov. 20 & DCT / IGRINS       & 50          & 600           & 0\farcs63 $\times$ 9\farcs3            \\
SP3                     & 2016 Nov. 20 & DCT / IGRINS       & 44.5        & 600           & 0\farcs63 $\times$ 9\farcs3     \\
SP4                     & 2016 Nov. 19 & DCT / IGRINS       & 90          & 300           & 0\farcs63 $\times$ 9\farcs3             \\
\enddata
\tablenotetext{a}{The spectral mapping on HH 167 with total 6 slit positions a$-$f.}
\tablenotetext{b}{The 2.7m Harlan J. Smith Telescope at the McDonald Observatory.}
\tablenotetext{c}{Total integration time at each slit position of the spectral map.}
\tablenotetext{d}{Deep pointing on slit position `b' in HH 167.}
\tablenotetext{e}{The 4.3m Discovery Channel Telescope at the Lowell Observatory.}
\tablecomments{The wavelength coverage and the resolving power of IGRINS are the same on HJST and DCT, which are 1.49$-$2.46 $\micron$ and $R$ $\equiv$ $\lambda/\Delta\lambda$ $\sim$ 45,000, respectively.}
\end{deluxetable}

\begin{deluxetable}{lcrrrr}
\tabletypesize{\scriptsize}
\tablecaption{Detected H$_{2}$ lines from SP1$-$4\tablenotemark{a}.\label{tbl:pdr}}
\tablewidth{0pt}
\tablehead{&&\multicolumn{4}{c}{Flux\tablenotemark{b}}\\
\cmidrule(rl){3-6}
Transition & $\lambda_{\rm vac}$ & \multicolumn{1}{c}{SP1}& \multicolumn{1}{c}{SP2}& \multicolumn{1}{c}{SP3}& \multicolumn{1}{c}{SP4}}
\startdata
4$-$2 O(3) & 1.509865                        &\nodata     &\nodata     &\nodata      & 25.69   $\pm$ 0.98 \\
5$-$3 Q(4) & 1.515792                        &\nodata     &\nodata     &\nodata      & 7.69    $\pm$ 0.92 \\
5$-$3 O(2) & 1.560730                        &\nodata     &\nodata     &\nodata      & 12.70   $\pm$ 0.69 \\
4$-$2 O(4) & 1.563516                        &\nodata     &\nodata     &\nodata      & 14.97   $\pm$ 0.62 \\
6$-$4 Q(1) & 1.601535                        &\nodata     &\nodata     &\nodata      & 20.57   $\pm$ 0.60 \\
5$-$3 O(3) & 1.613536                        &\nodata     &\nodata     &\nodata      & 21.96   $\pm$ 0.73 \\
6$-$4 Q(3) & 1.616211                        &\nodata     &\nodata     &\nodata      & 10.20   $\pm$ 0.54 \\
7$-$5 S(1) & 1.620530                        &\nodata     &\nodata     &\nodata      & 13.23   $\pm$ 0.58 \\
4$-$2 O(5) & 1.622292                        &\nodata     &\nodata     &\nodata      & 12.35   $\pm$ 0.71 \\
5$-$3 O(4) & 1.671821                        &\nodata     &\nodata     &\nodata      & 12.38   $\pm$ 0.68 \\
6$-$4 O(2) & 1.675019                        &\nodata     &\nodata     &\nodata      & 12.37   $\pm$ 0.54 \\
1$-$0 S(9) & 1.687721                        &\nodata     &\nodata     &\nodata      & 3.99    $\pm$ 0.65 \\
7$-$5 Q(1) & 1.728779                        &\nodata     &\nodata     &\nodata      & 12.42   $\pm$ 0.76 \\
6$-$4 O(3) & 1.732637                        &\nodata     &\nodata     &\nodata      & 18.52   $\pm$ 1.09 \\
5$-$3 O(5) & 1.735888                        &\nodata     &\nodata     &\nodata      & 8.63    $\pm$ 0.68 \\
1$-$0 S(7) & 1.748035                        &\nodata     &\nodata     &\nodata      & 1.99    $\pm$ 1.25 \\
1$-$0 S(6) & 1.787946                        &\nodata     &\nodata     &\nodata      & 0.50    $\pm$ 0.83 \\
1$-$0 S(3) & 1.957556                        & 81.30   $\pm$ 1.00 &\nodata     & 73.47   $\pm$ 3.39  & 69.22   $\pm$ 1.28 \\
1$-$0 S(2) & 2.033756                        & 32.32   $\pm$ 0.39 & 32.13   $\pm$ 1.04 & 31.86   $\pm$ 0.77  & 49.41   $\pm$ 0.56 \\
8$-$6 O(3) & 2.041816                        &\nodata     &\nodata     &\nodata      & 8.76    $\pm$ 0.62 \\
2$-$1 S(3) & 2.073510                        &\nodata     &\nodata     &\nodata      & 29.19   $\pm$ 1.87 \\
8$-$6 O(4) & 2.121570                        &\nodata     &\nodata     &\nodata      & 4.85    $\pm$ 0.43 \\
1$-$0 S(1) & 2.121831                        & 100.00  $\pm$ 0.39 & 100.00  $\pm$ 0.80 & 100.00  $\pm$ 0.65  & 100.00  $\pm$ 0.55 \\
2$-$1 S(2) & 2.154225                        & 4.51    $\pm$ 0.34 & 3.57    $\pm$ 0.96 & 3.96    $\pm$ 0.67  & 23.01   $\pm$ 0.48 \\
9$-$7 O(2) & 2.172704                        &\nodata     &\nodata     &\nodata      & 4.15    $\pm$ 0.53 \\
3$-$2 S(3) & 2.201397                        & 3.90    $\pm$ 0.36 &\nodata     & 3.75    $\pm$ 0.73  & 12.42   $\pm$ 0.54 \\
8$-$6 O(5) & 2.210741                        &\nodata     &\nodata     &\nodata      & 3.88    $\pm$ 0.90 \\
1$-$0 S(0) & 2.223299                        & 26.40   $\pm$ 0.38 & 21.25   $\pm$ 1.00 & 22.64   $\pm$ 0.74  & 62.32   $\pm$ 0.60 \\
2$-$1 S(1) & 2.247721                        & 12.96   $\pm$ 0.42 & 8.90    $\pm$ 1.04 & 13.42   $\pm$ 0.80  & 46.61   $\pm$ 0.67 \\
4$-$3 S(3) & 2.344479                        &\nodata     &\nodata     &\nodata      & 5.01    $\pm$ 0.67 \\
1$-$0 Q(1) & 2.406594                        & 112.94  $\pm$ 0.66 & 65.14   $\pm$ 5.38 & 98.34   $\pm$ 1.42  & 136.73  $\pm$ 0.96 \\
1$-$0 Q(2) & 2.413436                        & 36.00   $\pm$ 0.75 &\nodata     & 30.43   $\pm$ 1.58  & 71.53   $\pm$ 1.08 \\
1$-$0 Q(3) & 2.423731                        & 114.98  $\pm$ 0.81 & 86.02   $\pm$ 5.31 & 98.49   $\pm$ 2.06  & 76.71   $\pm$ 1.08 \\
1$-$0 Q(4) & 2.437491                        & 33.59   $\pm$ 0.71 & 23.80   $\pm$ 1.83 & 26.20  $\pm$ 1.58 & 31.83   $\pm$ 1.08\\
\enddata
\tablenotetext{a}{SP1$-$4 in Figure \ref{fig:slitposition}.}
\tablenotetext{b}{Reddening is corrected with $\Av$ $=$ 21, 16 and 7 for SP1, 2, and 3, respectively. In SP4, reddening is not corrected because $\Av$ $\sim$ 0. Fluxes are normalized to 1$-$0 S(1) line which is set to 100.}
\end{deluxetable}

\end{document}